\documentclass[superscriptaddress,aps,pra,english,floatfix,twocolumn,10pt]{revtex4-1}
\usepackage[T1]{fontenc}
\usepackage{amsmath}
\usepackage{textcomp}
\usepackage[latin1]{inputenc}
\usepackage{subfigure}
\usepackage{graphicx}
\usepackage{blindtext, rotating}
\usepackage{mathtools}
\usepackage{blindtext}
\usepackage{gensymb}
\usepackage{enumitem}
\usepackage{xcolor}
\usepackage{babel}
\usepackage{hyperref}
\usepackage{braket}
\usepackage{textcomp}
\usepackage{subfigure}  
\makeatother
\begin{document}

\title{Self-duality of One-dimensional Quasicrystals with Spin-Orbit Interaction}
\date{\today}
\author{Deepak Kumar Sahu}
\author{Aruna Prasad Acharya}
\affiliation{Department of Physics and Astronomy, National Institute of Technology, Rourkela, Odisha, India}
\author{Debajyoti Choudhuri}
\affiliation{Department of Mathematics, National Institute of Technology, Rourkela, Odisha, India}
\author{Sanjoy Datta} 
\email{dattas@nitrkl.ac.in}
\affiliation{Department of Physics and Astronomy, National Institute of Technology, Rourkela, Odisha, India}
\affiliation{Center for Nanomaterials, National Institute of Technology, Rourkela, Odisha, India}

\begin{abstract}
Non-interacting spinless electrons in one-dimensional quasicrystals, described by the Aubry-Andr\'{e}-Harper (AAH) 
Hamiltonian with nearest neighbour hopping, undergoes metal to insulator transition (MIT) at a critical 
strength of the quasi-periodic potential. This transition is related to the self-duality of the AAH Hamiltonian. 
Interestingly, at the critical point, which is also known as the self-dual point, all the single particle wave 
functions are multifractal or non-ergodic in nature, while they are ergodic and delocalized (localized)  below 
(above) the critical point. In this work, we have studied the one dimensional quasi-periodic AAH Hamiltonian in 
the presence of spin-orbit (SO) coupling of Rashba type, which introduces an additional spin conserving 
complex hopping and a spin-flip hopping. We have found that, although the self-dual nature of AAH Hamiltonian remains 
unaltered, the self-dual point gets affected significantly. Moreover, the effect of the complex and 
spin-flip hoppings are identical in nature. We have extended the idea of Kohn's localization tensor calculations for 
quasi-particles and detected the critical point very accurately. These calculations are followed by detailed 
multifractal analysis along with the computation of inverse participation ratio and von Neumann entropy, which 
clearly demonstrate that the quasi-particle eigenstates are indeed multifractal and non-ergodic at the critical 
point. Finally, we mapped out the phase diagram in the parameter space of quasi-periodic potential and SO coupling 
strength.      
\end{abstract}
\maketitle

\section{Introduction}
In one dimensional (1D) lattice with random disorders, all the electronic single-particle states (SPS) 
of the non-interacting Anderson Hamiltonian localize exponentially even if the strength of the disorder is  
arbitrarily small \cite{Anderson,Abrahams}. In pure Anderson Hamiltonian, metal (extended SPS) to insulator 
(localized SPS) transition exists only in three dimension (3D), while  
in two dimension (2D) there are no extended states, but for weak disorder SPS are marginally 
localized\cite{Abrahams}. The extended states are also ergodic, that is, in the thermodynamic limit the real 
space average of the $q$-th moment of $\left|\psi(i)\right|^2$ converges to its ensemble average value 
\cite{Deng-Sinha}. In contrast, a quasiperiodic lattice, described by the Aubry-Andr\'{e}-Harper (AAH) Hamiltonian 
\cite{Harper,Azbel,Aubry-Andre} with nearest-neighbour hopping, undergoes metal to insulator (MIT) transition at 
a critical disorder strength, even in 1D. The fundamental, spinless AAH Hamiltonian referred above can be represented
by the eigenvalue equation, $t (\psi_{n + 1} + \psi_{n-1}) + W \text{cos}(2\pi b n + \phi) \psi_{n} = E \psi_n$, where 
$W$ is the strength of the disorder, $t$ is the nearest neighbour hopping amplitude, and $\psi_n$ is the amplitude
of the electronic wave function at the lattice site $n$. When $b$ is irrational we have the 
quasi-periodic lattice. The AAH Hamiltonian in reciprocal space is given by,   
$\frac{W}{2} (\psi_{k + 1} + \psi_{k-1}) + 2 t \text{cos}(2\pi b n + \phi) \psi_{k} = E \psi_k$.
The Fourier transformed AAH Hamiltonian becomes same as the original Hamiltonian when
$W/t=2$ and their roles are interchanged. This unique feature of AAH Hamiltonian makes it 
\textit{self-dual}, as states below and above this critical value $W_c/t=2$ are related by Fourier
transformation. All the SPS are extended for $W_c/t < 2$ while they are localized for $W_c/t > 2$. The self-dual point 
$W_c/t=2$ is particularly special, as all the states show multifractal character, that is, they are extended but 
\textit{non-ergodic}. 

Disordered  electronic systems have continued to receive wide attention, even after 
more than five decades of the publication of the seminal paper by Anderson \cite{Anderson}. A major 
driving force behind this is due to the importance of disorders, which are unavoidable in real 
materials.  Also, the debate about whether MIT exists in dimension lower than three in such lattices 
is not settled completely. The search for MIT in lower dimensional disordered systems is 
not just an academic interest, but it is an extremely important problem from the point of view 
of device applications as well. Moreover, successful experimental simulations of these disordered 
Hamiltonians in optical lattice set-up \cite{roati,Lahini,Kraus,madsen} have opened the possibility 
of verification of the theoretical predictions. More recently, the critical behaviour of the spin-less 
version of the AAH Hamiltonian has been experimentally observed in polaritonic 1D wires, 
created with the help of cavity-polariton devices\cite{Zilberberg}. Consistent effort towards finding MIT in 
lower dimension have led researchers to consider Anderson Hamiltonian with SO coupling quite extensively 
\cite{Hikami, Evangelou-Ziman,Evangelou-PRL,Ando-PRB89,Janssen-Huckestein, Minakuchi,Yakubo,Asada-Slevin,Su-Wang}.
However, the effect of SO coupling in  quasi-periodic systems has not received much attention.  

In this present work, we have explored the effect of SO coupling on MIT in quasi-periodic 1D lattice. 
More specifically, we have studied the AAH model with nearest neighbour hopping in the presence of Rashba 
type spin-orbit (RSO) coupling \cite{Rashba}. RSO coupled electronic systems receive wide attention 
because of its potential in various device applications, especially in spintronic devices \cite{Zutic}. The 
main advantage of RSO coupled systems is the possibility of easier external control of the coupling strength. 
In order to find a comprehensive and clear answer, we have systematically used three different approaches. 
First, to determine the effect of RSO on the nature of MIT in 1D quasi-periodic lattice, we have used the idea 
of many-body localization tensor \cite{Resta-PRL98, Resta-Sorella-PRL99}, which is based on Kohn's idea of 
localization in disordered systems \cite{kohn}. In the process,  we have demonstrated that this many-body 
localization tensor, also known as Kohn's localization tensor(KLT), is a truly powerful method to study MIT 
for spinfull electronic systems, \textit{where spin states mix}. After obtaining precise answer on the 
effect of RSO on MIT, we studied the nature of the eigenstates across the entire energy spectrum for different 
disorder strengths. For this we have used the inverse participation ratio (IPR) and von Neumann entropy (vNE) 
to get a quick overall idea about the nature of the quasi-particle eigenstates. Finally, we have carried out 
detailed multifractal analysis to prove that the self-duality of AAH Hamiltonian remains 
preserved in the presence of RSO, although the self-dual point shifts towards higher disorder strength with the 
increase of RSO coupling. While carrying out the multifractral analysis, we have used both periodic and 
open boundary conditions (OBC) to show that OBC can be used for the quasi-periodic lattices as well, 
effectively removing the constraint on the system sizes for numerical computation of the multifractal 
spectrum. To summarize our conclusions, at the end, we present the phase diagram in the parameter space spanned 
by the disorder strength and RSO couplings. It is important to note that in an earlier study \cite{Kohomoto}, a 
1D quasi-periodic AAH Hamiltonian with RSO coupling was obtained starting from a tight-binding square lattice in 
presence of uniform magnetic field. However, the RSO Hamiltonian used in the above mentioned reference is not same 
as ours. Although, it had also been concluded that the self-dual nature remains intact, but it was reported that the
self-dual point remains unaffected by RSO coupling strength, while there exists two new phases, self-dual of each other,
in the parameter space of disorder strength and SO coupling. These phases are characterized by coexistence of 
delocalized and localized states. Naturally, we have not found any evidence of these phases. 

The paper is organized as follows: in Sec.\ref{Sec:localization-tensor}, we introduce briefly the
idea of Kohn's localization tensor (KLT) for non-interacting electrons. From the KLT calculations,
we can immediately identify the critical point for MIT. In Sec.~\ref{SubSec:Evolution-of-Critical-point},
we have explored the evolution of the critical point with the RSO coupling strengths. However, these 
results do not provide clear idea about the nature of the SPS at the critical point. To study the 
nature of the SPS, we have computed the single particle IPR spectrum in Sec.\ref{Sec:IPR} and the von Neumann 
entropy (vNE) in Sec.\ref{Sec:vNE}. 
To demonstrate that all the SPS are \textit{non-ergodic}
at the critical point, in Sec.\ref{Sec:MFS-Calculation}, we have calculated the multifractal 
spectrum for the three regions: below, above, and at the critical point. Finally, in 
Sec.\ref{Sec:Phase-Diagram}, we present the phase diagram with respect to the parameters that 
control the disorder strength and the strengths of RSO coupling. 
\section{Aubry-Andr\`{e} Model With RSO coupling}\label{Sec:Model}
The Hamiltonian considered in this work consists of two parts,
\begin{equation}
H'=H+H_R,
\label{Eq:eq1}
\end{equation}
where $H$ is the usual AAH Hamiltonian given by, 
\begin{eqnarray}
H &=& -t\displaystyle\sum_{{i=1},\sigma}^{L-1} (c^\dag_{{i+1},{\sigma}} c_{{i},{\sigma}}+ h.c) + \nonumber \\
 & & W\displaystyle\sum_{{i=1},{\sigma}}^{L}cos(2\pi bi+\phi)c^\dag_{{i},{\sigma}} c_{{i}.{\sigma}}
\label{Eq:eq2}
\end{eqnarray}
Here, $t$ is the hopping amplitude from site $i$ to site $i+1$ and $L = N a$ is the length of the lattice, where 
$N$ is the number of lattice sites and $a=1$(arbitrary unit) is the lattice spacing. 
$c^\dag_{{i},{\sigma}}$ and $c_{{i},{\sigma}}$ are the fermionic creation and annihilation operators respectively 
for spin operator $\sigma=\uparrow$,$\downarrow$ particle at site $i$. $W$ is the strength of quasi-periodic 
potential. $\phi$ is an arbitrary phase varying from $(0, 2\pi)$. The choice of the phase $\phi$ does not affect our 
conclusion, and henceforth we shall set it to zero. We have used $b = (\sqrt{5}+1)/2$. Please note that, 
sometimes in the literature $b = (\sqrt{5}-1)/2$ is also used, but our conclusions are independent of the 
particular choice of $b$. 

The RSO Hamiltonian $H_R$ is given by \cite{Birkholz},
\begin{eqnarray} 
H_R &=& -\alpha_z \displaystyle\sum_{{i=1},{\sigma},{\sigma'}}^{L-1} (c^\dag_{i+1,\sigma}(i\sigma_y)_{\sigma,\sigma'}
 c_{i,\sigma'}+ h.c.) \nonumber \\
 & & -\alpha_y \displaystyle\sum_{{i=1},{\sigma},{\sigma'}}^{L-1} (c^\dag_{i+1,\sigma}(i\sigma_z)_{\sigma,\sigma'}  
 c_{i,\sigma'}+ h.c.),
 \label{Eq:eq3}
\end{eqnarray}
where $\sigma_y$ and $\sigma_z$ are Pauli spin matrices in $y$- and $z$-direction respectively.
$\alpha_y$ is a complex spin-conserving hopping due to the confinement in $y$-direction and $\alpha_z$ 
is a spin-flip hopping due to the confinement in $z$-directions. The hopping amplitude 
$\alpha_y$ and $\alpha_z$ could be different in general and they could also be site dependent. 
The pure RSO Hamiltonian $H_R$, that is studied in this work, has also been studied in the context of 
transport properties in quantum nanowires \cite{Ando-Tamura,Mireles}. Also, recently 
localization properties of attractive fermions have been studied in presence of the spin-flip 
component of the RSO Hamiltonian\cite{Xianlong-PRA16}. Since $\alpha_{y}$ is a spin 
preserving hopping process, it is expected that it will not change the self-dual nature of AAH 
Hamiltonian. However, it is not clear how $W_c/t$ is going to be affected by $\alpha_y$ alone. 
On the other hand, the effect of spin-flip hopping, $\alpha_z$, on the self-duality as well 
as on $W_c/t$ is not evident. Hence, we are going to focus mainly on the problem where $\alpha_z \neq 0$ 
and $\alpha_y =0$. However, for completeness, we also present KLT results for $\alpha_y \neq 0$ and 
$\alpha_z = 0$. 

\begin{figure*}[ht]
\centering
	\includegraphics[width=0.32\textwidth,height=0.25\textwidth]{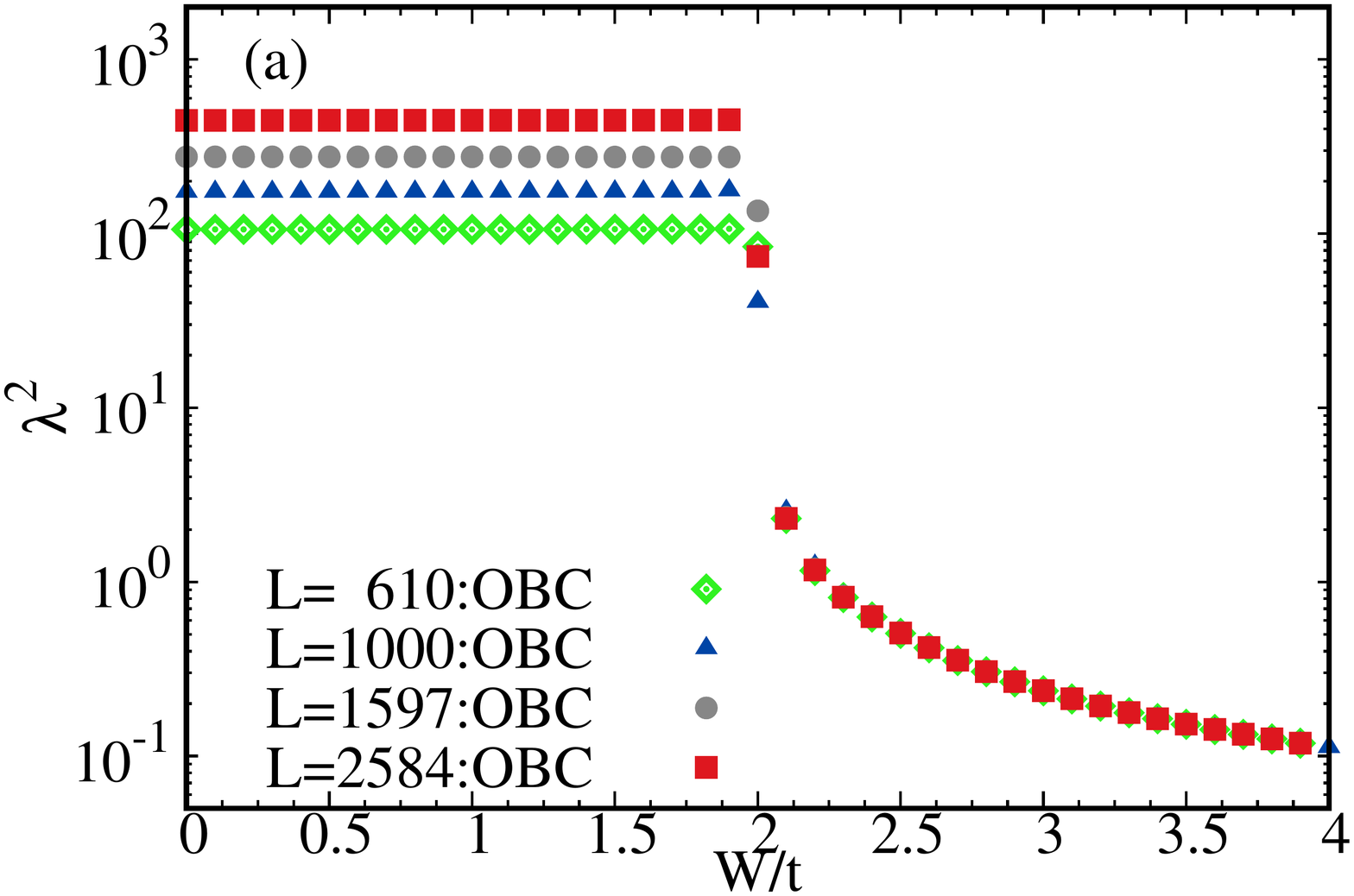}\hspace{-0.5cm}
	\includegraphics[width=0.32\textwidth,height=0.25\textwidth]{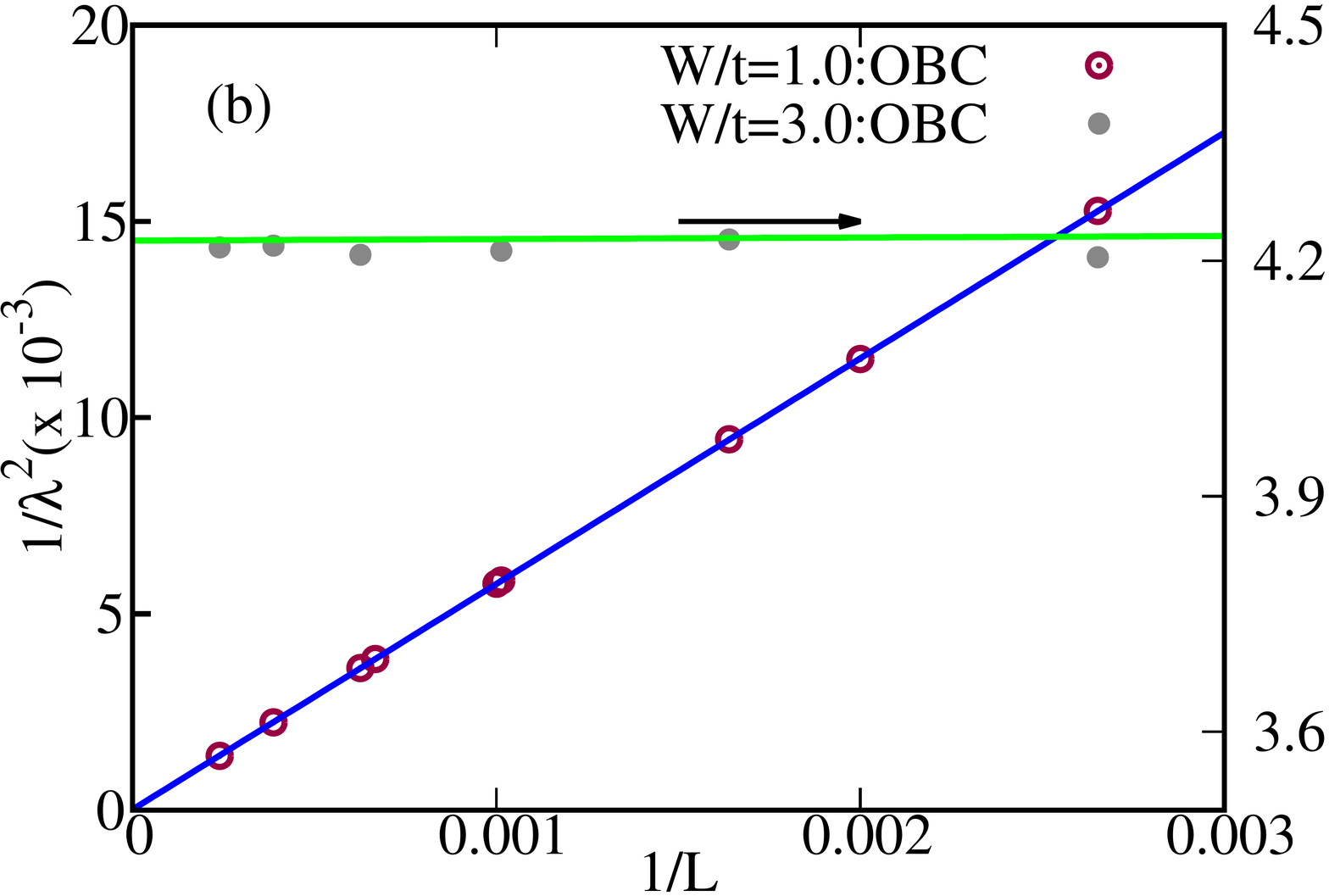} \hspace{-0.5cm}
	\includegraphics[width=0.32\textwidth,height=0.25\textwidth]{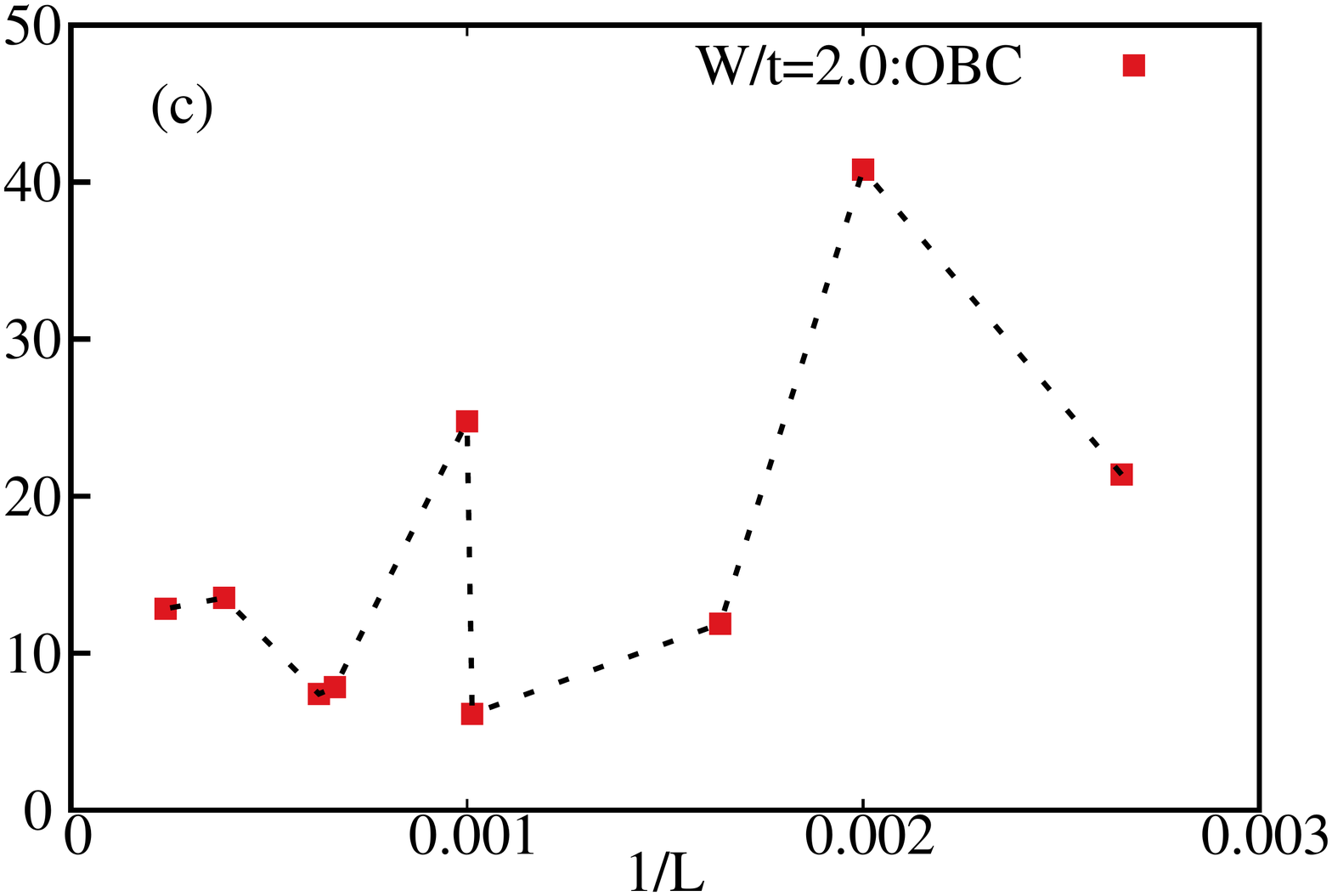}
\caption{(a) Squared localization length $\lambda^2$ for half-filled 1D pure AA model 
with respect to disorder strength $W/t$ for some selected lattices. (b) Scaling of inverse localization 
length $1/\lambda^2$ with inverse chain length $(1/L)$. For finite size scaling we have used 
$1/\lambda^2 = a_0 + b_0/L$. (c) Plot of $1/\lambda^2$ with $1/L$ at the critical point.
It is evident that at the critical point, $1/\lambda^2$ does not follow the scaling pattern of 
either metallic or insulating states. All the results for open boundary condition.}
\label{Fig:localization-obc}
\end{figure*}
\begin{figure*}[ht]
\centering
	\includegraphics[width=0.32\textwidth,height=0.25\textwidth]{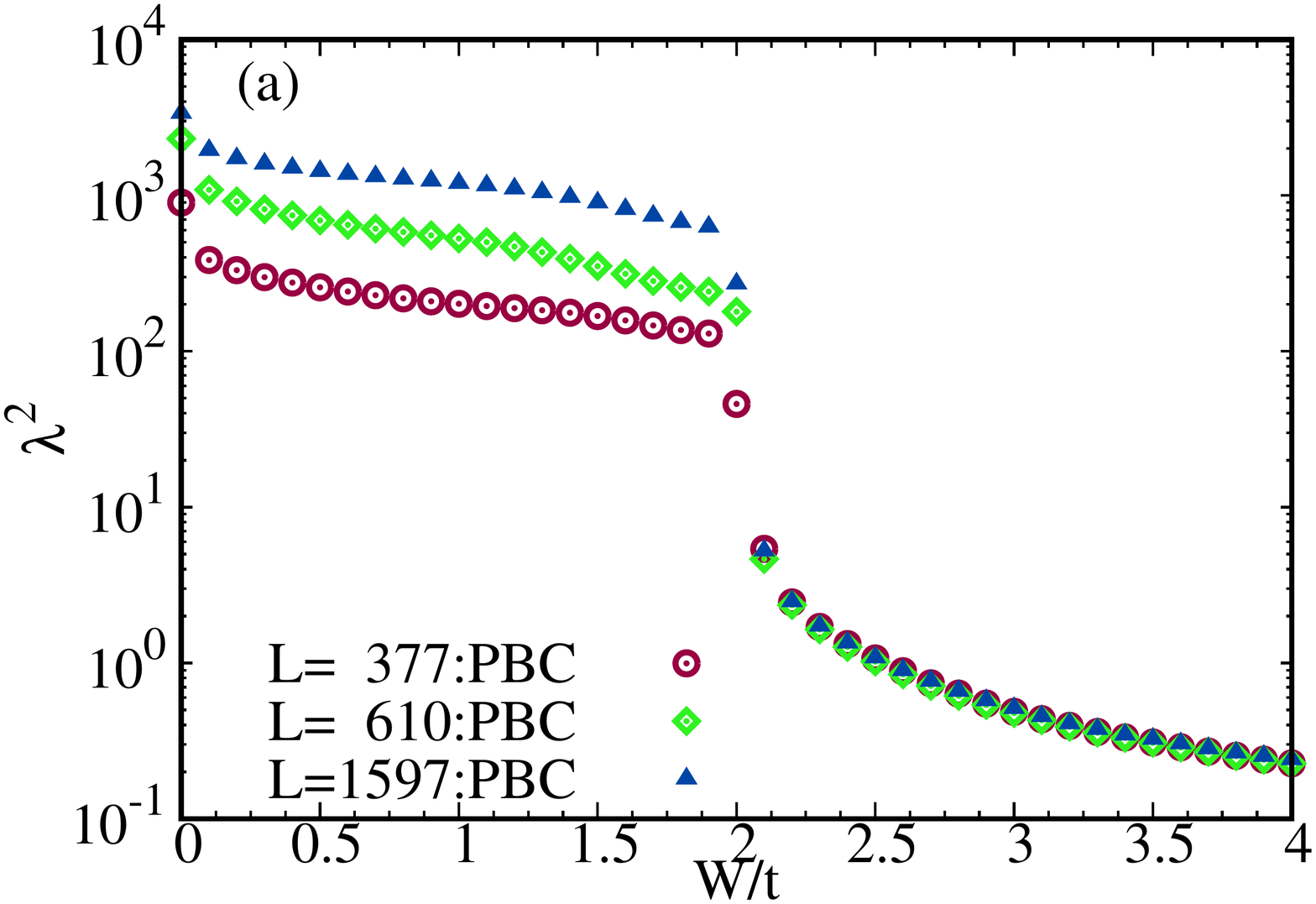}\hspace{-0.5cm}
	\includegraphics[width=0.32\textwidth,height=0.25\textwidth]{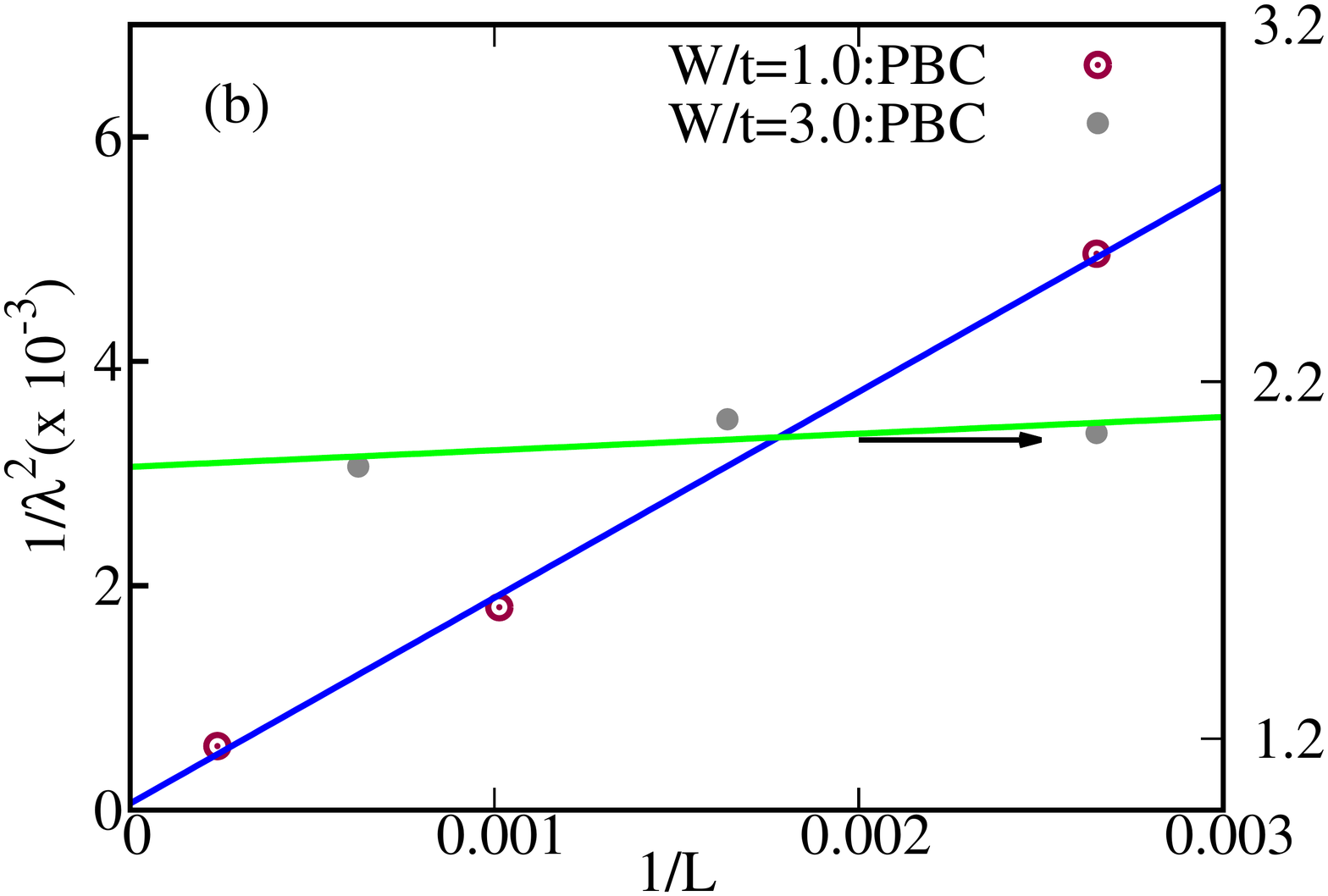} \hspace{-0.5cm}
	\includegraphics[width=0.34\textwidth,height=0.25\textwidth]{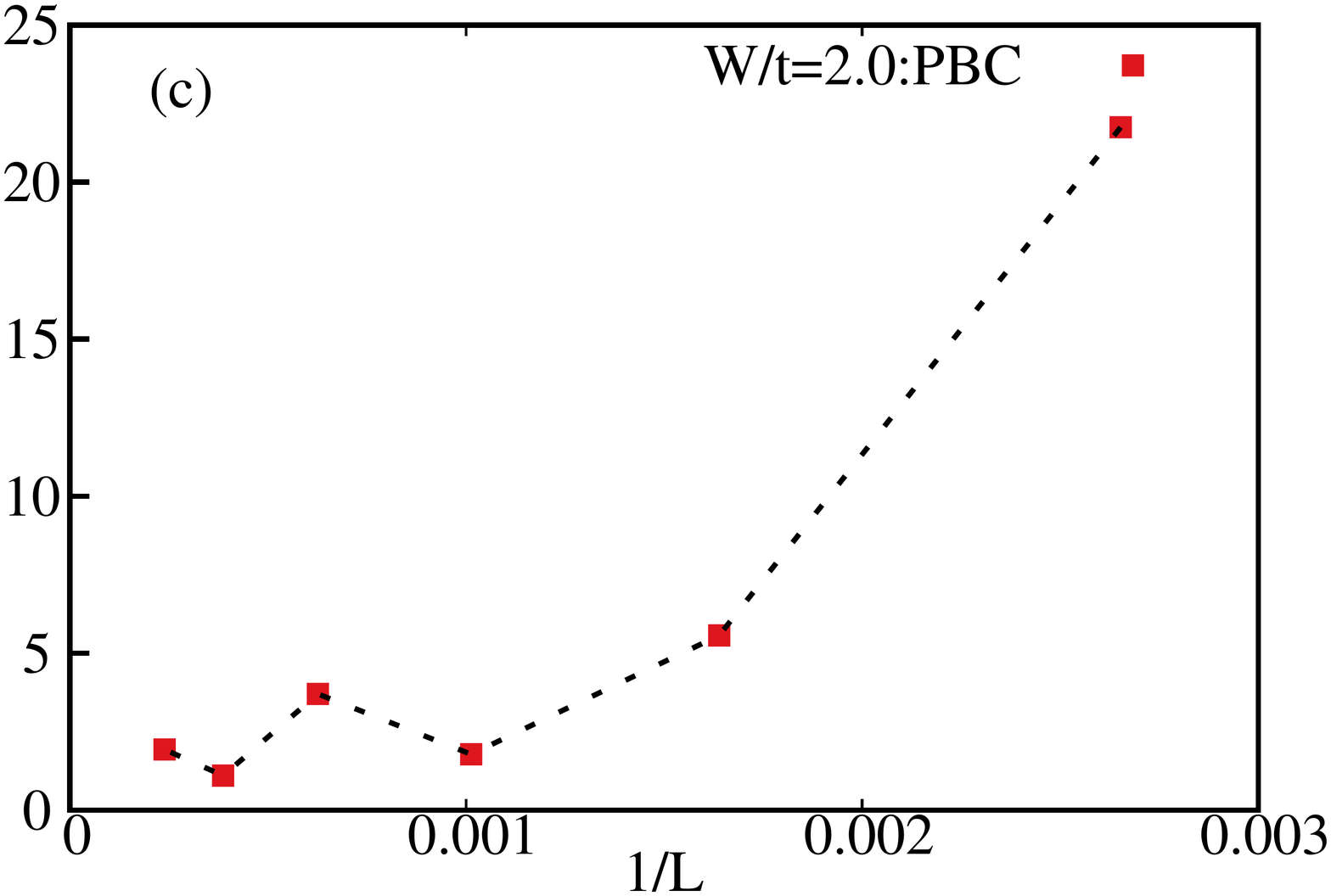} 
\caption{Localization tensor $\lambda^2$ for half-filled 1D pure AAH model with PBC. 
(a) $\lambda^2$ vs. $W/t$ for a few selected chain lengths. (b) Finite size scaling of $1/\lambda^2$ vs. $(1/L)$.
(c) Once again, similar to the OBC case, $1/\lambda^2$ lacks simple scaling behaviour with $1/L$ at the 
critical point.}	
\label{Fig:localization-pbc}
 \end{figure*}
\section{Kohn's Localization Tensor and Metal to Insulator Transition}
\label{Sec:localization-tensor}
For a comprehensive analysis of the interplay of disorder and RSO coupling
we have systematically used different approaches to capture the full picture.  
In this section, we discuss the idea of Kohn's localization tensor. 
It is a reliable way to characterize metallic and insulating state. Based on the idea first proposed by 
W. Kohn\cite{kohn}, Resta and Sorella \cite{Resta-PRL98, Resta-Sorella-PRL99} formulated a localization tensor, 
alternatively called as Kohn's localization tensor (KLT). In the thermodynamic limit, it is independent of 
system size for insulating states, while for metallic states it diverges. For 1D AAH model with nearest-neighbour
hopping and no RSO coupling, a metal to insulator transition takes place at the critical disorder strength $W_c/t=2.0$. 
Recently, it has been shown \cite{varma} that the localization tensor $\lambda_{\alpha\beta}$, where $\alpha$ 
and $\beta$ are the space co-ordinates, can capture this transition accurately. Furthermore, it has been shown 
that computation of  $\lambda_{\alpha\beta}$ becomes particularly simpler for non-interacting electrons as well as 
for interacting electrons, if interaction is treated within mean-field approximation. However, in these two 
above mentioned scenarios spin states were completely decoupled. As mentioned earlier, in the presence of spin-flip 
hopping induced by RSO coupling, spin states mix to giving rise to quasi-particle eigenstates. Here, in this 
work we extended the computational approach of $\lambda_{\alpha\beta}$ for quasi-particle states and 
simultaneously capture the MIT, if it exists, accurately. We have presented our results for periodic and 
open boundary conditions. However, as we are going to see, we have found that it is not always possible to 
calculate $\lambda_{\alpha\beta}$ with open boundary condition.  

Here, we briefly discuss the key aspects of KLT and methods to calculate it for different 
boundary conditions \cite{Resta-JPC-02-review,varma,Resta-EurPhys2011}. A more elaborate discussion on KLT and 
ways to compute it can be found in Ref.\cite{Resta-JPC-02-review} and 
in Ref.\cite{Resta-JCP06}. For periodic boundary condition (PBC), the localization tensor, 
$\lambda_{\alpha\beta} (\text{here~} \alpha= x,\beta =x)$ can be expressed as,
\begin{equation}
 \lambda_{xx}^2=-\frac{L^2}{4\pi^2N}ln\frac{|z_{N}^{x}||z_{N}^{x}|}{|z_{N}^{xx}|},
 \label{Eq:Lambda_def}
\end{equation}
where the quantity $z_N^{x}$ is given by,
\begin{equation}
z_{N}^{x}=\langle\Psi\rvert e^{i\frac{2\pi}{L}\hat{\mathbf{R}}_{x}} \lvert\Psi\rangle,
\label{Eq:Z_def}
\end{equation}
whereas, $z_{N}^{\alpha \beta}$ can be obtained by replacing $\hat{R}_{\alpha}$ 
with $\hat{R}_{\alpha}-\hat{R}_{\beta}$. Hence, in our case $|z_{N}^{xx}|=1$. 
In the above equation, $\ket{\Psi}$ is the many-body ground state wave function, $\hat{\mathbf{R}}=\sum_{i=1}^{N} \hat{r}_i$ is the the many-body position operator with $\hat{\mathbf{R}}_{x}$ being the $x$ component. $N$ is the number of lattice sites, while $L = N~a$, with $a=1$(in arbitrary unit) being the lattice constant,  is the size of the system. For a half-filled system ($L=N$) in 1D, the localization tensor reduces to 
\begin{equation}
\lambda_{xx}^2=\lambda^2=-\frac{L}{2\pi^2} \text{ln} \left\{|z_N|\right\}.
\label{Eq:Z_def_half_fill}
\end{equation}

In absence of electron-electron interaction, $z_N^{(x)}$ can be simplified 
further \cite{Resta-JPC-02-review,Resta-EurPhys2011} and represented as $z_N^{(x)}={det}^2[S_{jj'}^{x}]$, 
where $S_{jj'}^{x}$ is a matrix whose elements are  given by,
\begin{equation}
S_{j,j'}^{x}=\int dr~\psi_{j}^*(r) e^{i\frac{2\pi}{L}\hat{r}_{x}}\psi_{j'}(r).
\label{Eq:S_jj'_def}
\end{equation}
In the above equation $\psi_{j}(r)$ represents the amplitude of single-particle wave function at position $r$ 
for a spin-up or spin-down electron arranged in ascending order in energies. The indices $j$ and $j'$ indicate the energy level. Since, in absence of RSO spin-up and spin-down electrons are completely decoupled, it is sufficient to consider only one type of spin and $j,j'=1,2,\cdots,N/2$ to compute $\lambda_{xx}^2$ at half-filling. 
But, in presence of RSO, more specifically because of the spin-flip hopping process, in Eq.~\ref{Eq:S_jj'_def}  
$\phi_{j}(r)$ now represents the amplitude of a single quasi-particle wave function at position $r$ 
corresponding to the $j^{th}$ eigen-energy. Hence, in our case $j,j'=1,2,\cdots,N.$ 

In case of open boundary conditions (OBC), squared localization length $(\lambda^2)$, in units of the nearest-neighbour distance, can be expressed as follows \cite{Resta-JCP06,Bendazzoli-Resta-JCP2010}:
\begin{equation}
\lambda^2=\frac{1}{\nu N}\displaystyle\sum_{i,i'=1}^{N}\rho_{ii'}^2(\nu)(i-i')^2,
\label{Eq:Lambda^2_OBC_def}
\end{equation}
where $i,i'=1,\cdots,N$ represent the lattice site, $\nu$ is the filling factor and $\rho_{ii'}(\nu)$ is the 
one-body density matrix, defined as
\begin{equation}
  \rho_{ii'}=\displaystyle\sum_{j=1}^{N}\phi_j(i)\phi_j^*(i'),
  \label{Eq:one-body-density-matrix-OBC}
\end{equation}
where $\phi_j(i)$ is the amplitude of single quasi-particle wave function at lattice site $i$ corresponding to
$j^{th}$ eigenvalue. Since we are interested to compute $\lambda^2$ at half-filling, we set $\nu=1$ in 
Eq.~\ref{Eq:Lambda^2_OBC_def} .  
\begin{figure*}[ht]
\centering
	\includegraphics[width=0.32\textwidth,height=0.234\textwidth]{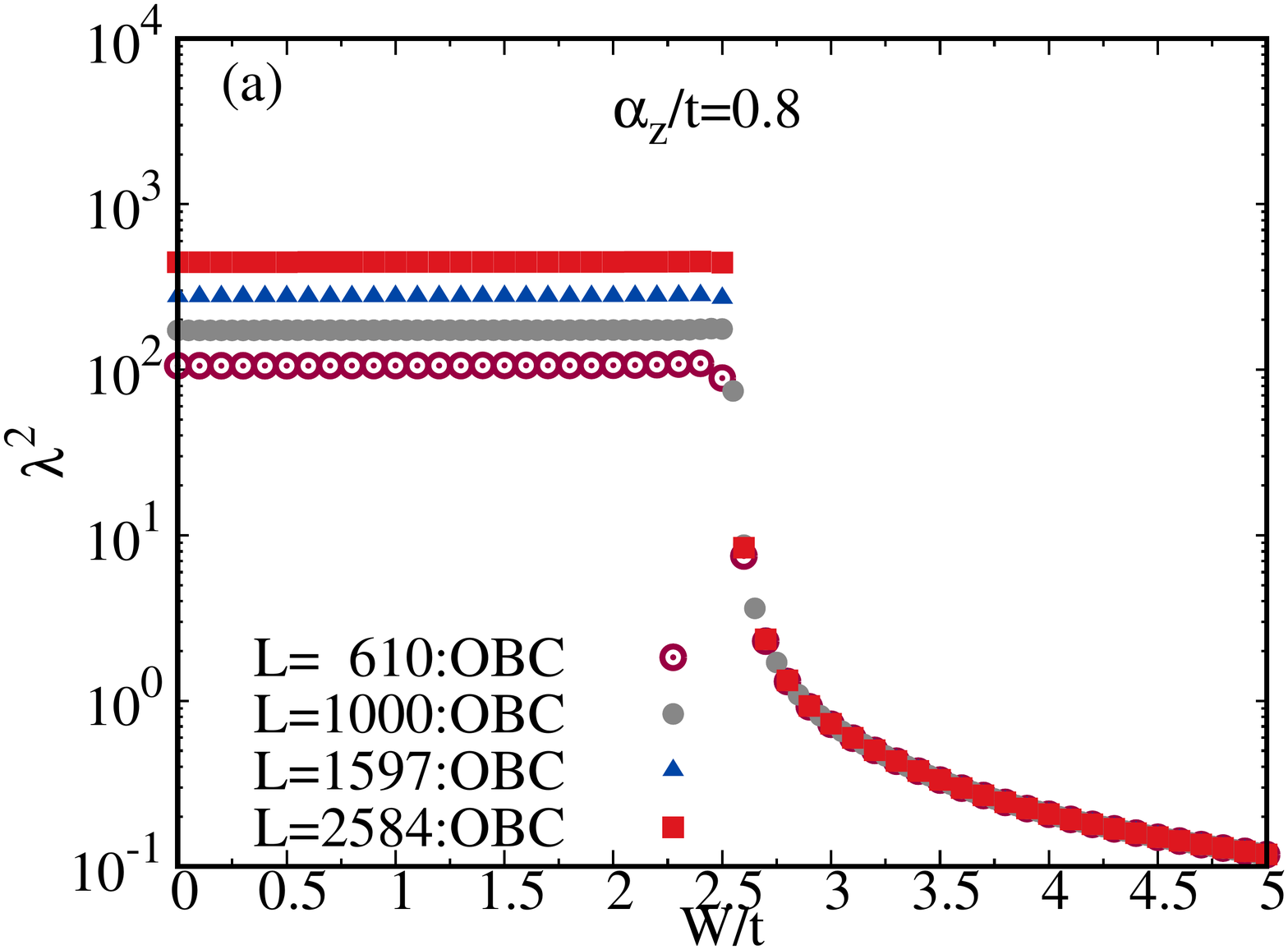}\hspace{-0.5cm}
	\includegraphics[width=0.32\textwidth,height=0.25\textwidth]{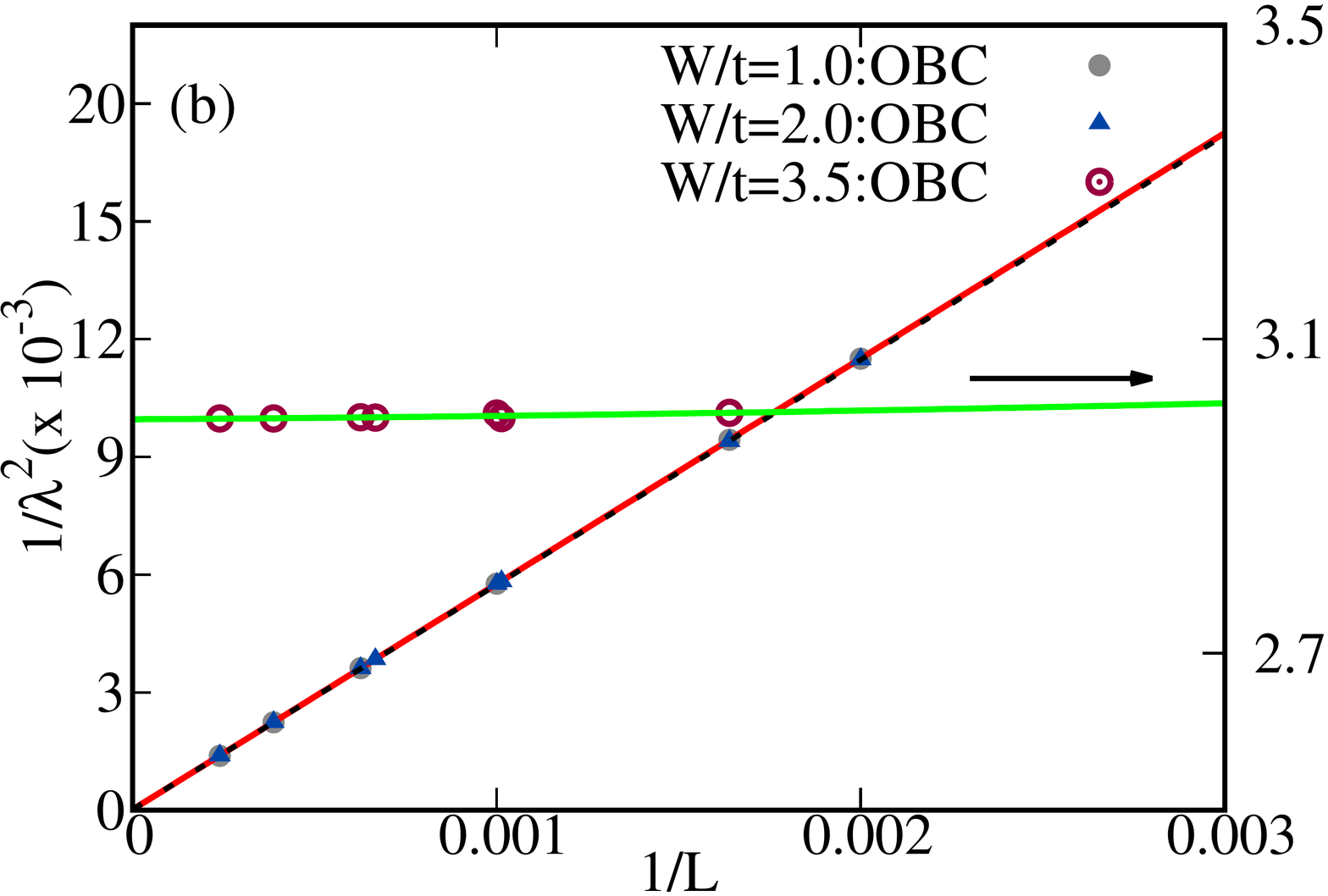} \hspace{-0.5cm}
	\includegraphics[width=0.32\textwidth,height=0.25\textwidth]{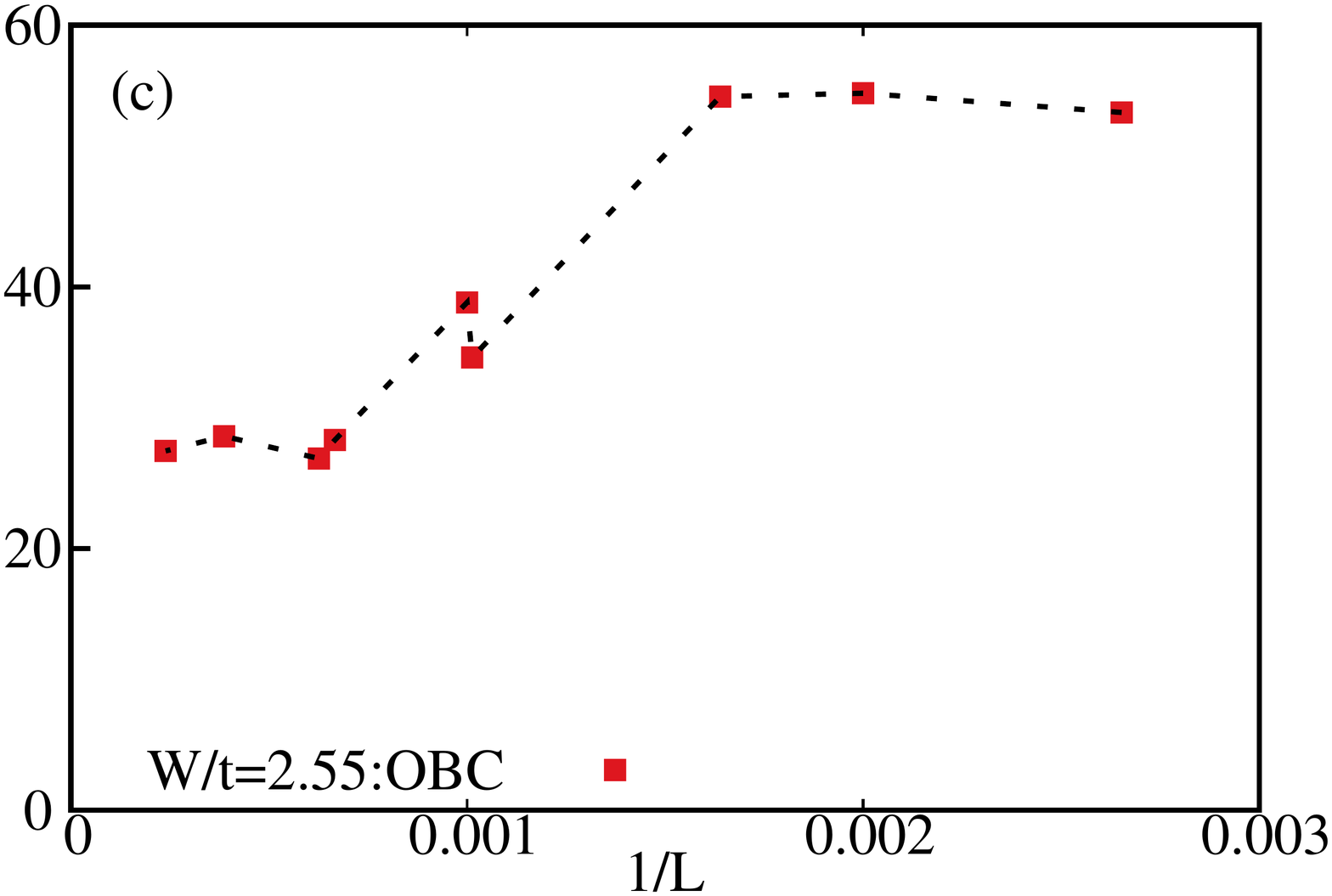} 
\caption{Localization tensor $\lambda^2$ for half-filled 1D AAH model with RSO coupling. (a)$\lambda^2$ vs. $W/t$ 
for selected lattices with OBC. (b) Scaling of inverse localization length $1/\lambda^2$ vs. $1/L$. Here $1/\lambda^2$
scales linearly with $1/L$ similar to Fig.~\ref{Fig:localization-obc}(b). (c) Plot of $1/\lambda^2$ vs. $1/L$
at the critical point $W_c/t = 2.55$. Lack of proper scaling of $1/\lambda^2$ indicates that our estimation of critical 
point is quite correct.} 
 	\label{Fig:localization-so-obc}
 \end{figure*}
 \begin{figure*}
\centering
	\includegraphics[width=0.32\textwidth,height=0.234\textwidth]{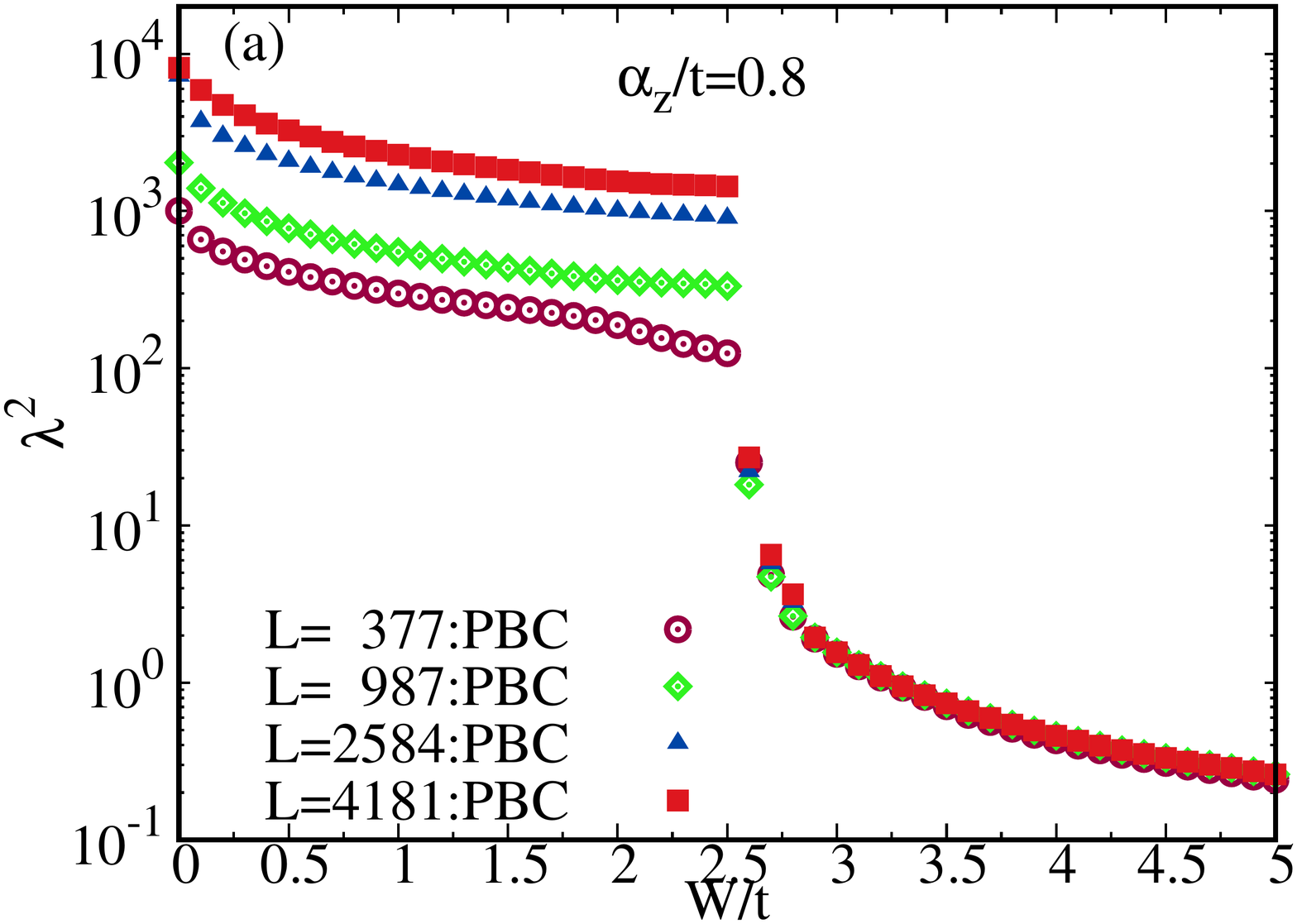}\hspace{-0.5cm}
	\includegraphics[width=0.32\textwidth,height=0.25\textwidth]{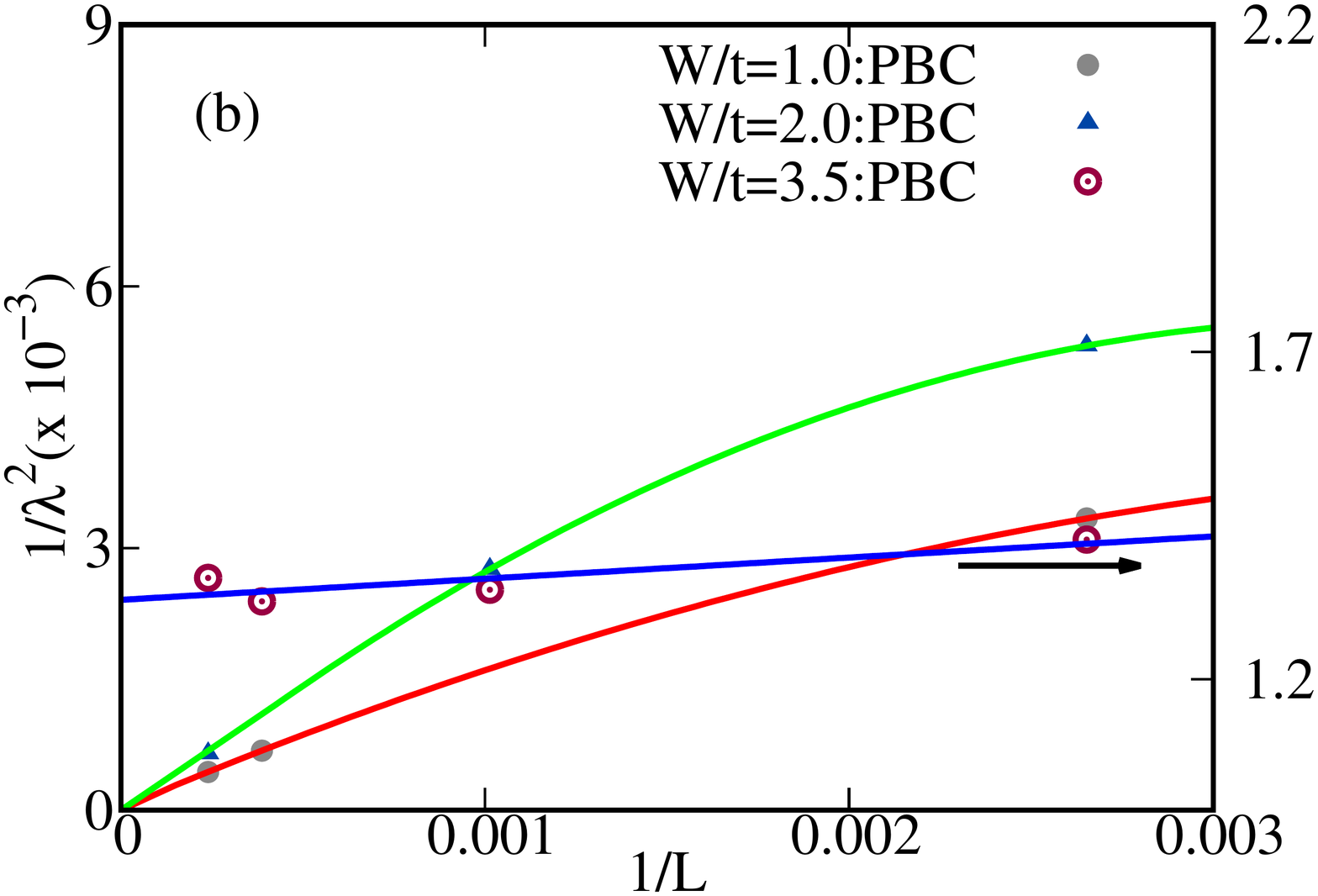}\hspace{-0.5cm}
	\includegraphics[width=0.32\textwidth,height=0.25\textwidth]{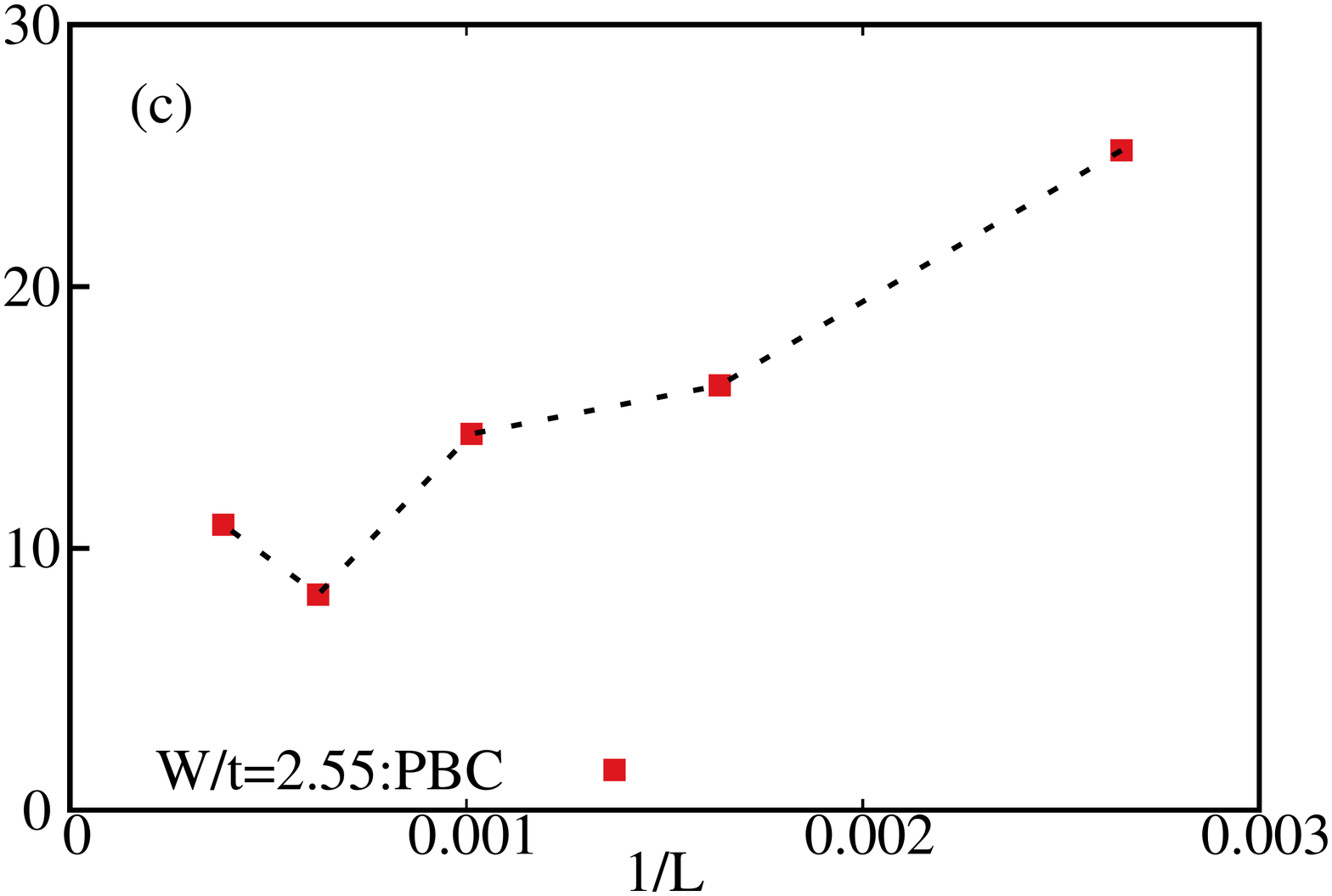}
\caption{Localization tensor $\lambda^2$ for half-filled 1D AAH model with RSO coupling. (a) $\lambda^2$ vs. $W/t$ 
for selected lattices with PBC. (b) Finite size scaling of $1/\lambda^2$ with inverse chain length $1/L$ below and above the critical point. We have used $1/\lambda^2 = a_0 + b_0/L + c_0/L^2$ for the purpose of scaling in the metallic phase. (c) With PBC as well we find anomalous behaviour of $1/\lambda^2$ with increasing system size at $W/t=2.55$.}
\label{Fig:localization-so-pbc}
\end{figure*}
\subsection{Localization tensor without RSO}{\label{SubSec:loc-tensor-wo-RSO}}
Before presenting the localization tensor results for our Hamiltonian, in this section we first benchmark our $\lambda^2$ calculations 
for pure AA model without the RSO coupling at half-filling. We also highlight the behaviour of $\lambda^2$ with system sizes at the 
critical point. In Fig.\ref{Fig:localization-obc}(a) and Fig.\ref{Fig:localization-pbc}(b),
we have shown these calculation for open and periodic boundary conditions respectively. In case of PBC, system sizes are restricted to lattice sizes given by the Fibonacci sequence, while for OBC there is no such restriction. One of the main objectives to compare results of different boundary conditions is to show that the localization tensor is a robust indicator to differentiate metallic and insulating state irrespective of the boundary conditions. This observation is going to be useful, as we are going to show in Sec.\ref{Sec:multifractal-spectrum}, while computing the multifractal spectrum for quasi-periodic lattices, as we do not need to restrict ourselves to very specific system sizes dictated by the Fibonacci sequence in order to extrapolate the results to thermodynamic limit.

In Fig.\ref{Fig:localization-obc}(a)(OBC) and Fig.\ref{Fig:localization-pbc}(a)(PBC), we have shown the variation of 
squared localization length $\lambda^2$ with increasing disorder strength $(W/t)$ for different 1D lattices at half-filling. 
As expected, the transition from delocalized to localized phase occurs at $W_c/t\simeq2$ for both the boundary conditions. 
$\lambda^2$ values are finite and independent of system sizes above $W/t>2$ whereas for $W/t<2$, it increases with 
the increase of system size. These observation are similar to Ref.~\cite{varma}.  
In Sec.\ref{Sec:localization-tensor} it has been mentioned that in case of metallic phase $\lambda^2$ diverges
in the thermodynamic limit, while it is independent of system sizes in the insulating phase. From the plot of $1/\lambda^2$ vs. $1/L$
of Fig.~\ref{Fig:localization-obc}(b) and Fig.\ref{Fig:localization-pbc}(b), it is clear that, irrespective of the boundary
condition, $\lambda^2$ diverges for $L \rightarrow \infty$ in the metallic phase, while it is nearly constant and converges
to a finite value in the insulating phase. For pure AA Hamiltonian, we have found that $1/\lambda^2$ scales linearly with the 
inverse of the system size. For finite size scaling, in this case, we have used $1/\lambda^2= a_0 + b_0/L$, where $a_0$ and 
$b_0$ are two adjustable parameters. 

At the critical point $W_c/t=2$, scaling behaviour of $\lambda^2$ with increasing system size is expected 
to be erratic as all the single particle eigenstates are mutifractral in case of pure AA model. 
However, it can be an useful and quick indicator to detect the existence of multifractal eigenstates within a spectrum. 
Since there is no characteristic length-scale for multifractal states, we can expect an anomalous behaviour of 
$1/\lambda^2$ with respect to $1/L$ compared to pure metallic and insulating phases. In Fig.~\ref{Fig:localization-obc}(c) and 
in Fig.~\ref{Fig:localization-pbc}(c) we have plotted $1/\lambda^2$ versus $1/L$ for $W_c/t=2$. In case of OBC, it is clear that 
$\lambda^2$ neither converges to a finite value nor does it go to zero in the thermodynamic limit. With PBC as well, 
$\lambda^2$ oscillates with increasing $L$ without any indication of convergence. In this work, however, we have used 
a well established and computationally less costly approach for the multifractal analysis of the eigenstates.

\subsection{Localization tensor in presence RSO}{\label{SubSec:loc-tensor-w-RSO}}
We now present our localization tensor results for one-dimensional AA model in the presence of RSO coupling.   
Since the effect of the spin-flip hopping process, induced by RSO, 
on the self-duality is much harder to anticipate, we primarily focus on the $\alpha_y = 0$ case.   
We have chosen $\alpha_z/t=0.8$. There is nothing special about this value, as all our conclusions are independent of it, except the location of self-dual point. In Fig.\ref{Fig:localization-so-obc}(a)(OBC) and \ref{Fig:localization-so-pbc}(a)(PBC) the variation of squared localization length, $\lambda^2$, with increasing value of disorder strength $W/t$ have been shown. It is evident that inclusion of RSO coupling enhances the critical point towards larger disorder strength, in this case $W_c/t\simeq2.5$, as compared to pure AA model ($W_c/t\simeq2$). Furthermore, this conclusion is independent of the applied boundary conditions on the system.  

To confirm our observation in the thermodynamic limit we have done finite size scaling of  
$1/\lambda^2$ with different system sizes below and above the critical point. These results are 
presented in Fig.~\ref{Fig:localization-so-obc}(b) and in Fig.~\ref{Fig:localization-so-pbc}(b). 
In the case of OBC, it is evident that $1/\lambda^2$ converges to zero as $L\rightarrow \infty$ for $W/t < 2.5$, that is, for delocalized phase, while it is independent of system size for $W/t > 2.5$, indicating an insulating phase. 
In case of PBC also, these fundamental conclusions remain same. Interestingly, however, with PBC the dependence of $1/\lambda^2$ on $1/L$ deviates significantly in the presence of RSO in the metallic phase. For finite size scaling, specially for the metallic phase, we have used $1/\lambda^2= a_0 + b_0/L + c_0/L^2$, where $a_0, b_0$ and $c_0$ are adjustable parameters.

It is easy to pinpoint the critical disorder strength from the $\lambda^2$ vs $W/t$ data almost exactly. We are going to see that this information can reliably be extracted from the multifractal analysis (Sec.~\ref{Sec:MFS-Calculation}) as well. From the localization tensor result  we estimate $W_c/t \simeq 2.55$ for $\alpha_z/t=0.8$. To cross-check this estimation, we have plotted $1/\lambda^2$ versus $1/L$ separately at $W_c/t \simeq 2.55$ in 
Fig.~\ref{Fig:localization-so-obc}(c) and in Fig.~\ref{Fig:localization-so-pbc}(c).
It is quite clear that $1/\lambda^2$ does not follow any discernible pattern with $1/L$. This, as we have 
pointed out in Sec.~\ref{SubSec:loc-tensor-wo-RSO}, is the hallmark of critical/multifractal eigenstates. 
\begin{figure*}[ht]
\centering
    \includegraphics[width=0.32\textwidth,height=0.25\textwidth]{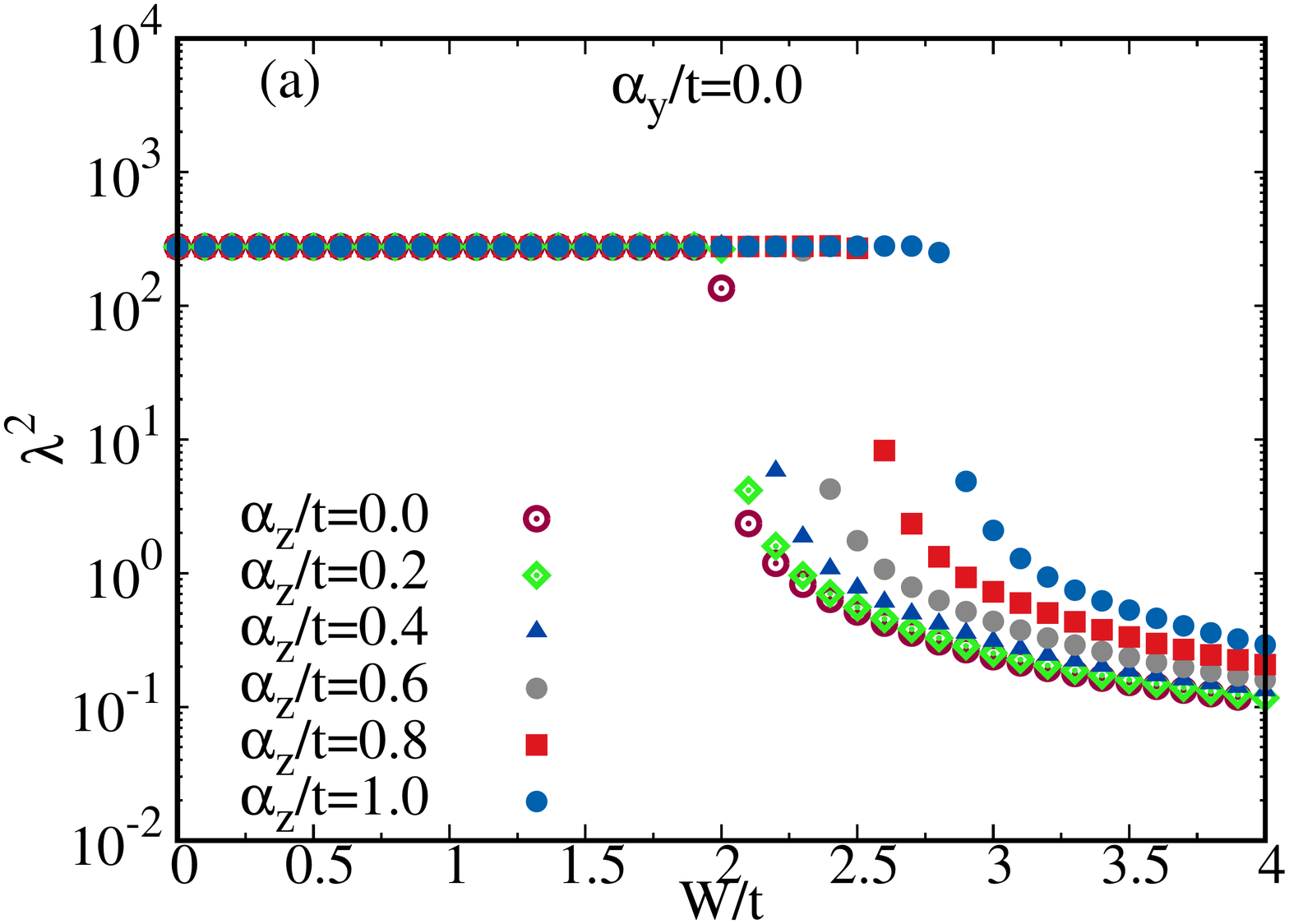}
    \includegraphics[width=0.32\textwidth,height=0.25\textwidth]{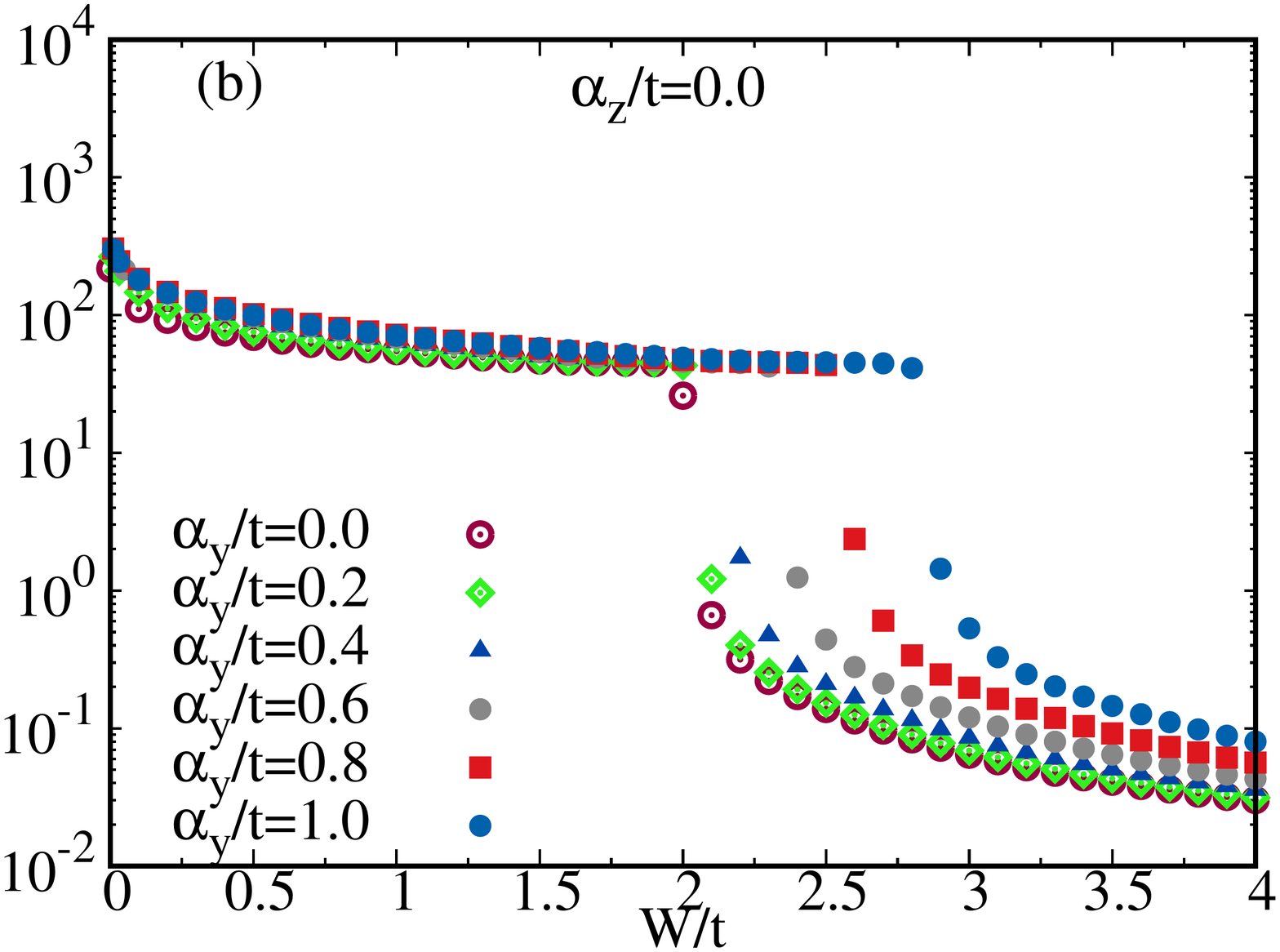}
    \includegraphics[width=0.32\textwidth,height=0.25\textwidth]{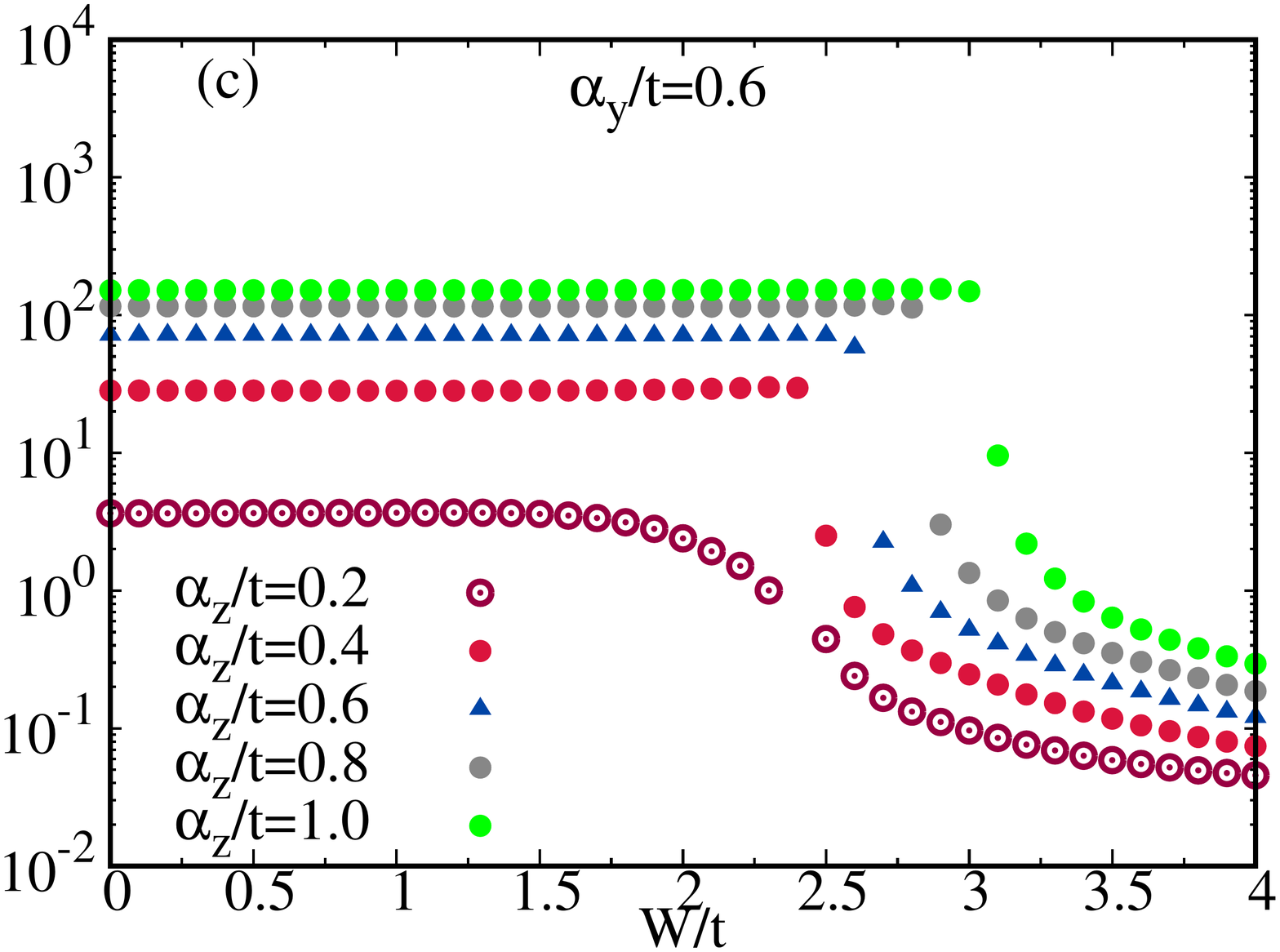}
\caption{(a) Evolution of the self-dual point in AAH with only spin-flip hopping induced by RSO coupling.  
The results are for a lattice $L=1597$ with OBC. (b) The self dual point moves towards higher strength 
of the quasi-periodic potential when pure AAH Hamiltonian is considered with only the complex hopping 
induced by the RSO coupling. (c) Evolution of the self-dual point when the AAH Hamiltonian is considered
with the full RSO Hamiltonian. Here we have only shown the change with the spin-flip hopping while
the complex hopping has been fixed at a value $0.6$.}
\label{Fig:localization-tensor-vs-alphaz-n-y}
\end{figure*}
\subsection{Evolution of the critical point with RSO interaction}{\label{SubSec:Evolution-of-Critical-point}}
In this section, we study the evolution of the critical point for three different cases, (i) $\alpha_y = 0$, (ii) $\alpha_z=0$
and (iii) $\alpha_z \neq 0, \alpha_y = \rm{fixed}.$  For case (i) and (iii) a lattice with $N=1597$ and OBC have been used, while for case (ii) we have used PBC and $N=987$. It is important to note that $\lambda^2$ could not be computed with OBC for case (ii). In Fig.~\ref{Fig:localization-tensor-vs-alphaz-n-y}(a), we have considered only the spin-flip hopping $\alpha_z$ in the RSO Hamiltonian. It is clear, even from the results for a single finite lattice, that the critical point moves to higher strength of the quasi-periodic potential. Interestingly, when  only the complex hopping of the RSO Hamiltonian is considered along with pure AA Hamiltonian, similar variation of the critical point is observed in Fig.~\ref{Fig:localization-tensor-vs-alphaz-n-y}(b).   
Fig~\ref{Fig:localization-tensor-vs-alphaz-n-y}(c), represents the general nature of the evolution of the critical point when the full RSO Hamiltonian is considered along with pure AA Hamiltonian. In these results, variation of the critical point with spin-hopping has been studied for a fixed strength of the spin conserving complex hopping having the amplitude $\alpha_y/t =0.6$.     
For $\alpha_z=0.2$, $\lambda^2$ appears to indicate a cross-over rather than a clear transition. However, this is an artifact of the finite system being used for these calculations. We have checked that as the system size is increased,  a clear transition from delocalized to localized transition emerges. 
\section{Inverse Participation Ratio}{\label{Sec:IPR}}
Existence of MIT in presence of RSO is evident from the results of the localization 
tensor calculations. Although at the critical point, anomalous behaviour of the localization tensor
with the inverse of the system size do indicate the absence of a characteristic length scale, the true nature of the quasi-particle eigenstates across the entire energy spectrum is not clear. For example, if an energy spectrum contains predominantly delocalized states then the localization tensor can suppress the contribution of localized states while calculating it upto certain filling fraction. To distinguish localized and delocalized states, Inverse Participation Ratio (IPR) is regularly used as a first indicator. IPR can also provide some hint, although only qualitatively, of multifractal eigenstates, if it exists. Generally, IPR is used to quickly identify the existence of mobility-edge. Typically the mobility edge is defined as the energy which separates localized and delocalized eigenstates in the energy spectrum. 
If a mobility edge exists, IPR jumps from system size independent (insulating/localized states) higher value to a lower value (ergodic metallic/extended states) that scales inversely with the system size. 
In our case, we have observed that away from the critical point, the states are either delocalized $(W < W_c)$ or localized $(W > W_c)$. The other important question that remains to be answered is, what are the nature of the eigenstates at the critical point? In AA model without RSO, all the eigenstates are extended but non-ergodic at the critical point. IPR data qualitatively indicates that RSO does not alter this behaviour. 
At this point we would like to mention that, studying the single particle energy spectra $E_n$ and the distribution of the level-spacing $\delta_n = E_{n+1} -E_{n}$ can also reveal a great deal about the nature of the eigenstates \cite{Evangelou-Pichard,Takada,Machida}; for example, in AAH model the energy level-spacing follows Poissonian distribution, while at the critical point this distribution follows an 
inverse power law. Typically, the energy spectrum of AAH Hamiltonian consists of subbands and many gaps between these subbands. Recently, it has been reported that there are some special states inside these subband gaps \cite{Roy-Sharma}, which are localized even in the metallic phase of pure AAH Hamiltonian. We have checked that these states are mostly concentrated at the lattice edges. Appearance of these states depends on many factors; for example, they can appear for system with OBC, with PBC, but the system size is not same as certain (not all) Fibonacci number. These states do not influence the critical behaviour in any way, as can be seen from the results of the previous section. 
\begin{figure*}[ht]
\centering
\includegraphics[width=0.32\textwidth,height=0.25\textwidth]{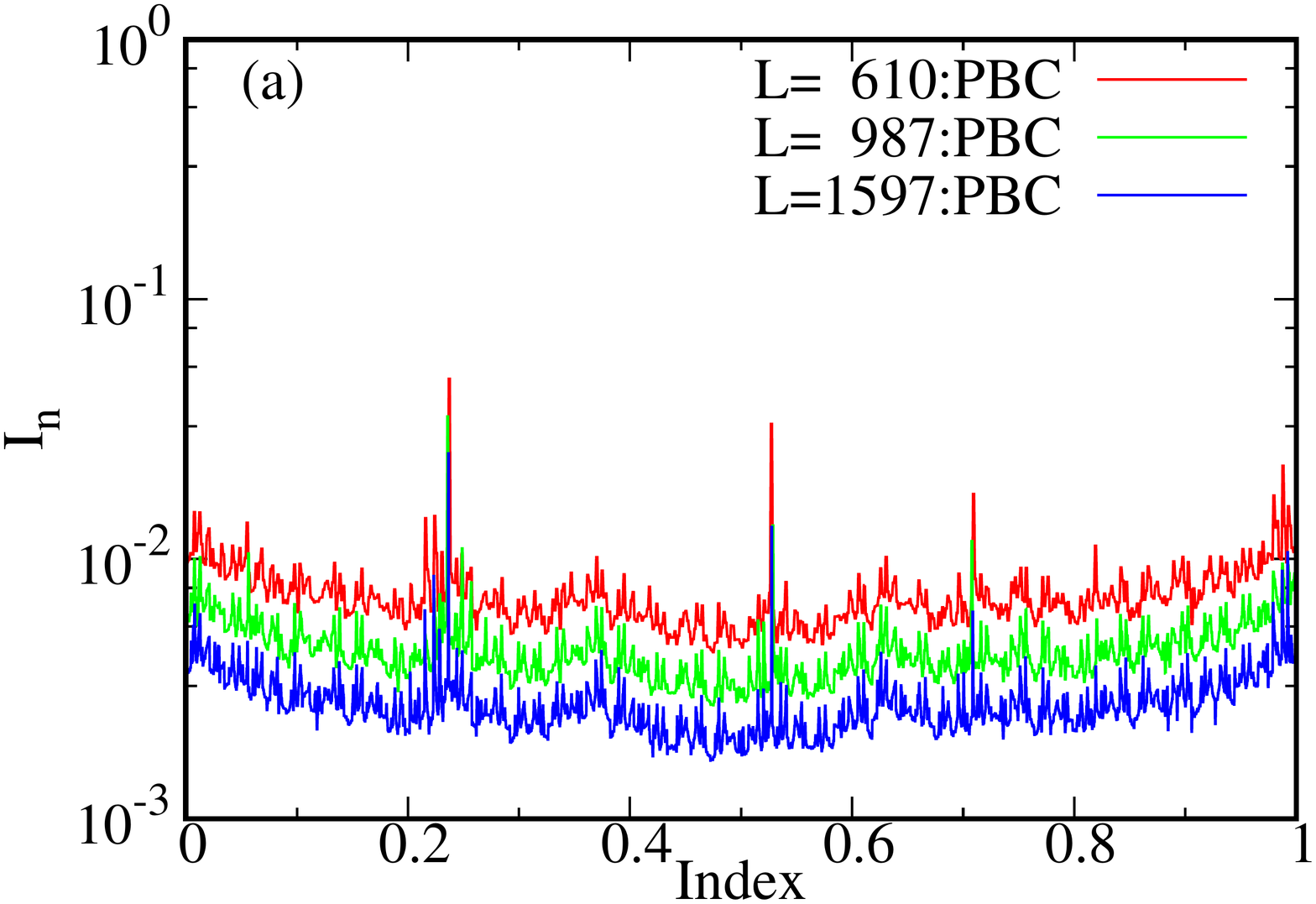}
\includegraphics[width=0.32\textwidth,height=0.25\textwidth]{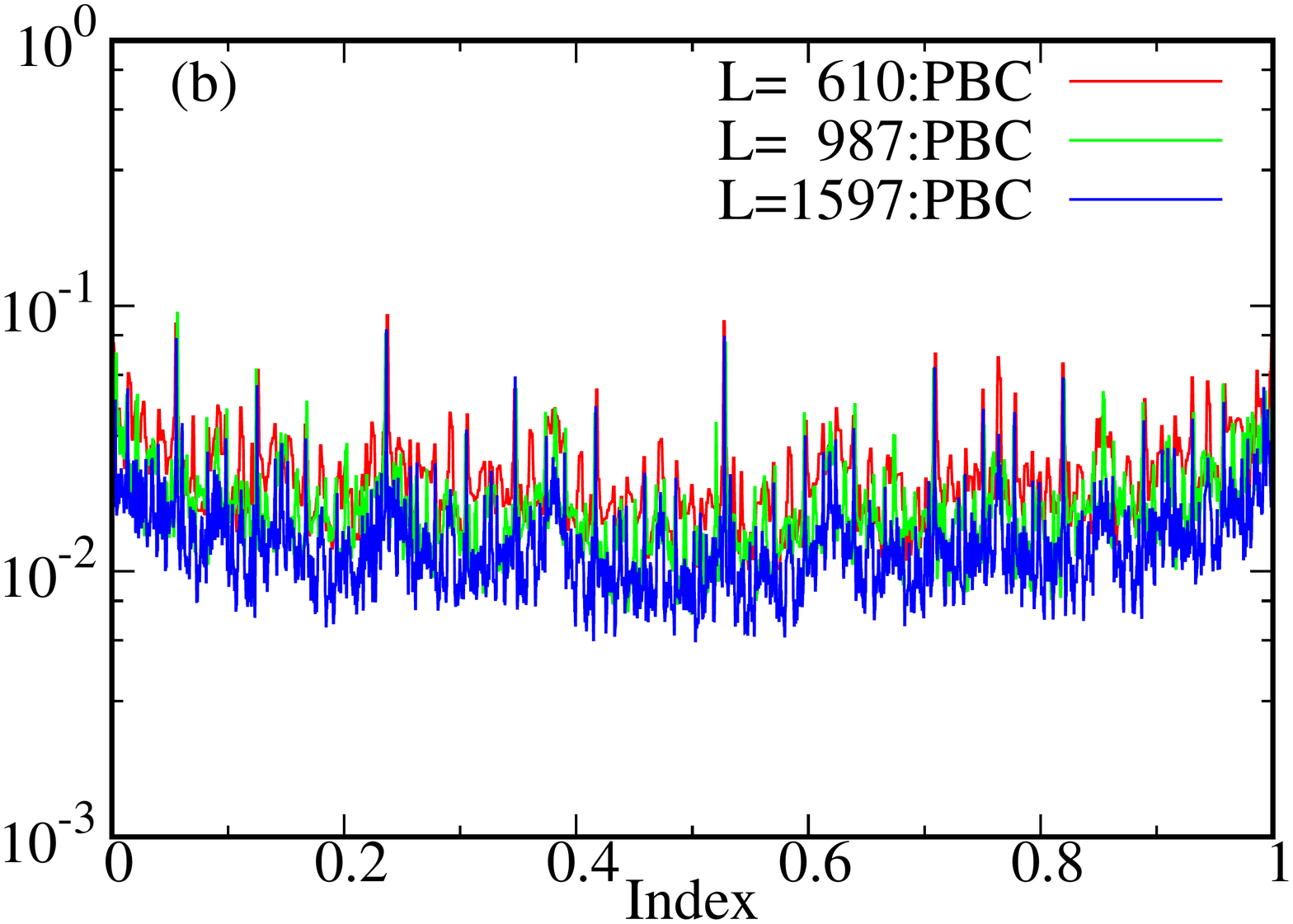}
\includegraphics[width=0.32\textwidth,height=0.25\textwidth]{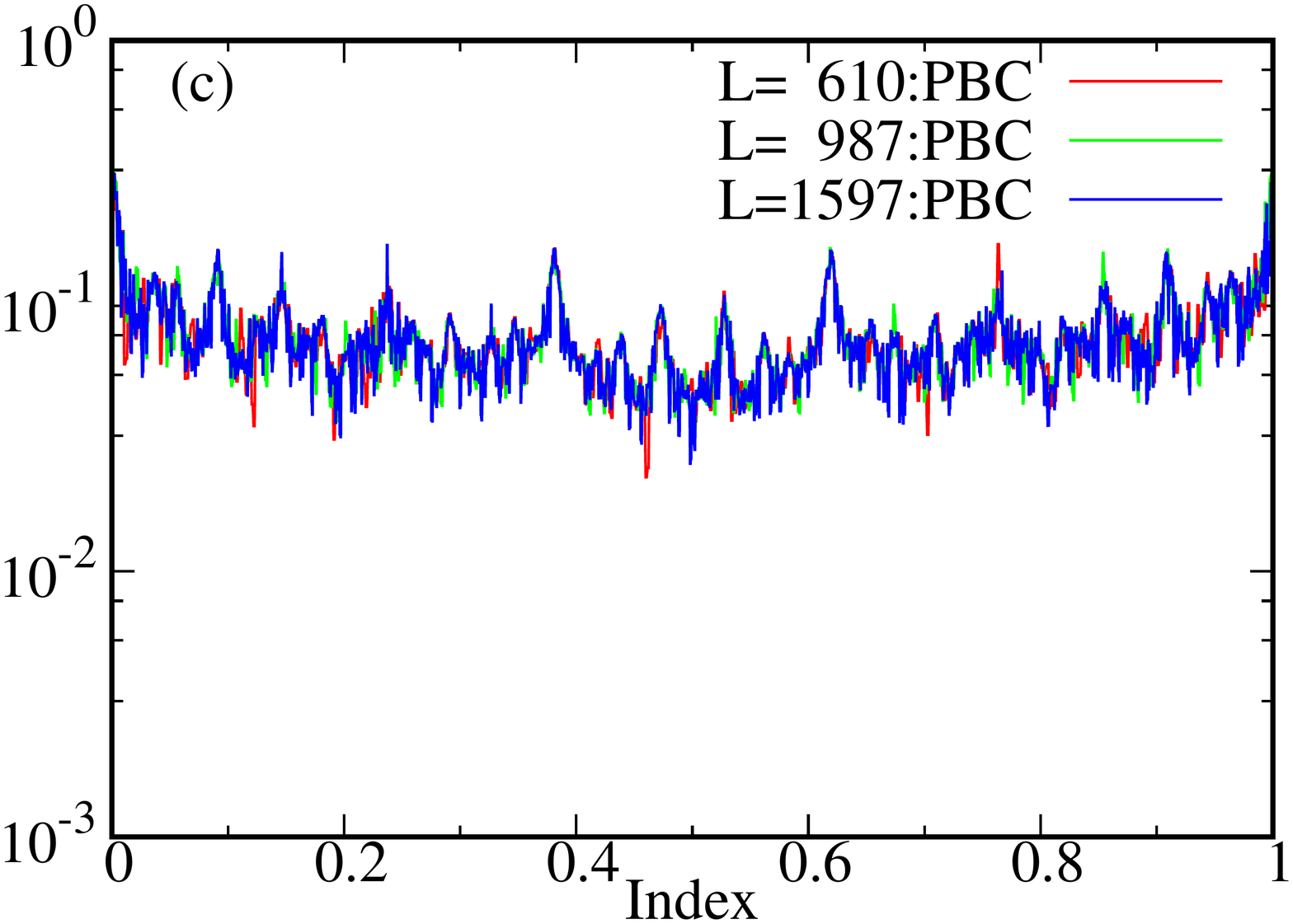}
\caption{Evolution of IPR ${I_n}$ with increasing disorder strength for a fixed RSO coupling
$\alpha_z/t=0.8$ and $\alpha_y=0.$ (a)$W/t=2.5$,(b)$W/t=2.55$, and (c) $W/t=2.6$. Here index represents the ratio of serial number of eigenstate to total number of eigenstates ($\mathrm i/(2L)$), where $i=1,2,\cdots,2L$.}
\label{Fig:IPR}
\end{figure*}
\begin{figure*}
\centering
\includegraphics[width=0.32\textwidth,height=0.25\textwidth]{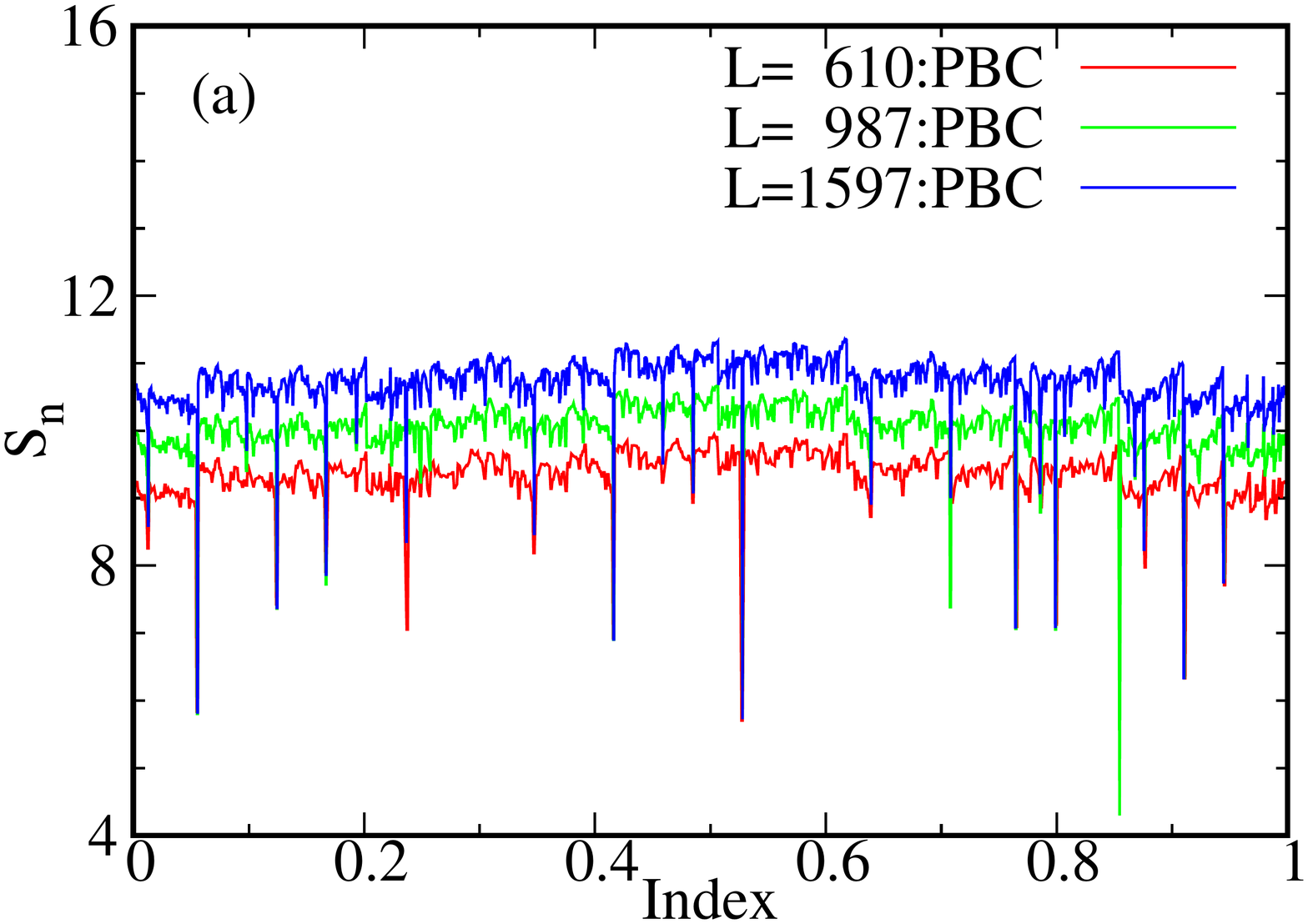}
\includegraphics[width=0.32\textwidth,height=0.25\textwidth]{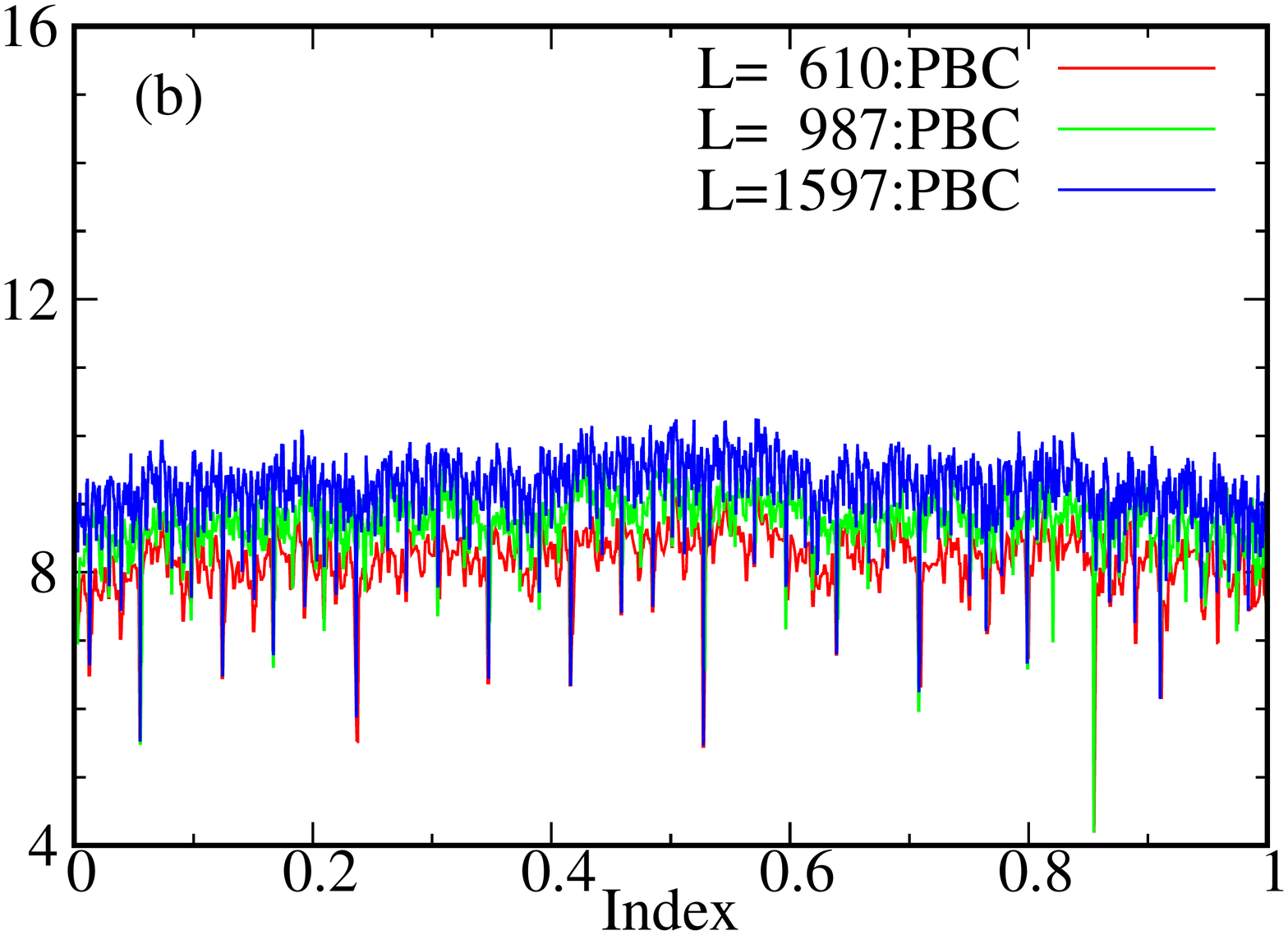}
\includegraphics[width=0.32\textwidth,height=0.25\textwidth]{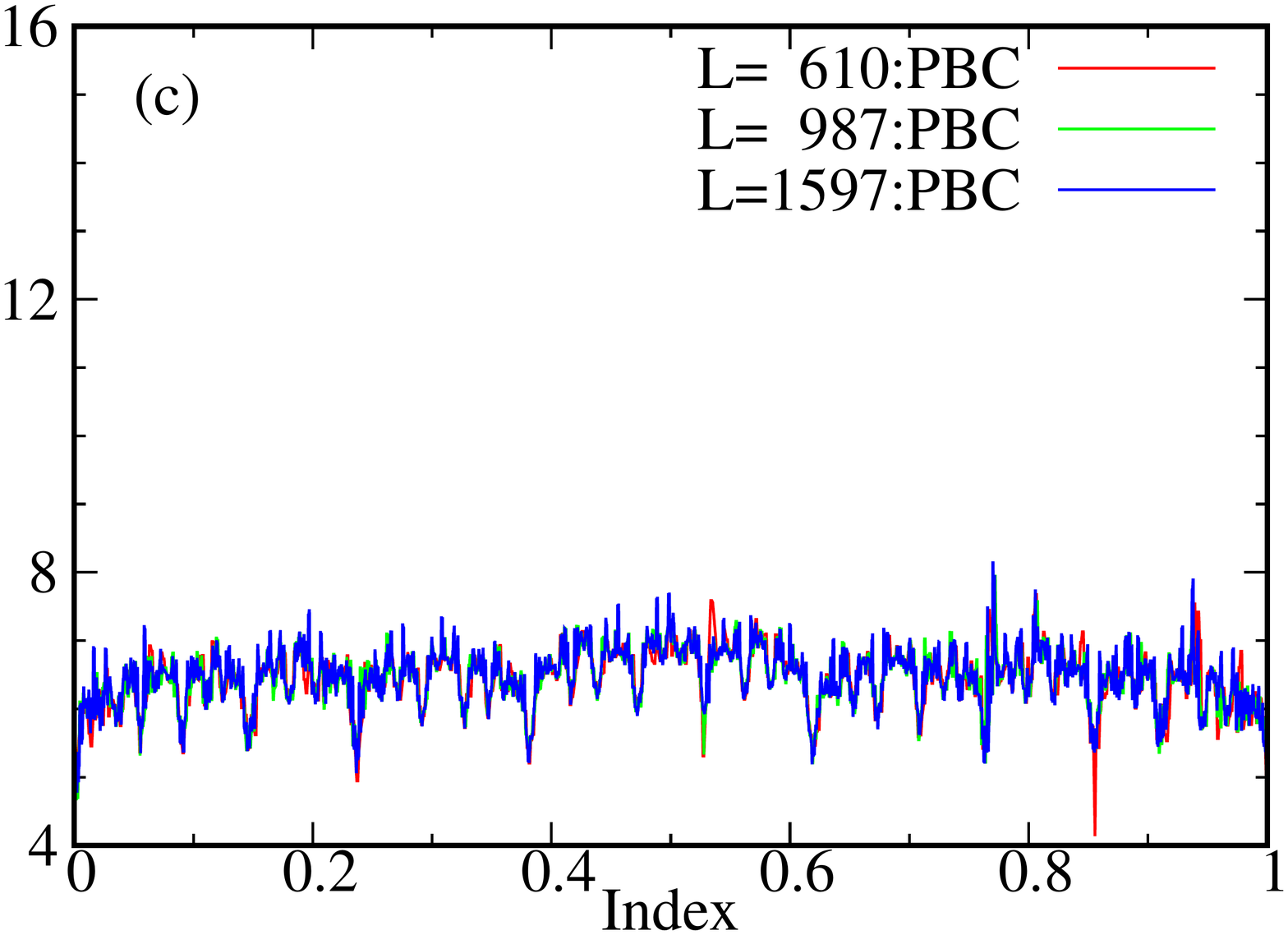}
\caption{Evolution of von Neumann Entropy ${S_n}$ with increasing disorder strength for a fixed RSO coupling
$\alpha_z/t=0.8$ and $\alpha_y=0.$ (a)$W/t=2.5$,(b)$W/t=2.55$, and (c) $W/t=2.6$.}	
\label{Fig:Entropy}
\end{figure*}

Formally, for $\nu$-th quasiparticle eigenstate the usual definition of IPR \cite{Evers-Mirlin} can be generalized as 
follows;
\begin{equation}
P_{\nu}(q) =\sum_{\sigma} \sum_{i=1}^{N} \left|\psi_{\nu,\sigma}(i) \right|^{2 q},~~q=2.  
\end{equation}
Here $\nu$th single quasiparticle eigenstate is given by 
$\ket{\Psi_{\nu}} = \sum_{\sigma} \sum_{i=1}^{N} \psi_{\nu,\sigma}(i) \ket{1,\sigma}_{i},$
where $\ket{1,\sigma}_{i} = \ket{0,0,\cdots,\sigma_i,\cdots,0,0}$ represents the localized 
basis state having one particle with spin $\sigma$ at site $i$. Since, this is a non-interacting problem,
usual scaling properties with respect to system size is also expected to hold for the quasiparticle states, i.e. 
for a perfectly extended  metallic state $P_{\nu}(q) = 1/N$, and for a completely localized state $P_{\nu}(q)=1$. 
For a localized state, the IPR value is supposed to be system size independent. These distinct scaling properties of IPR enables it to identify delocalized and localized states quickly. We have numerically verified that the spin-up and spin-down contributions are identical towards the IPR of the quasiparticle they constitute.

In Fig.~\ref{Fig:IPR} we have plotted the IPR spectrum for (a) $W/t=2.5$, 
(b) $ W/t=2.55$, and (c)$W/t= 2.6$. These results are for $\alpha_z/t=0.8, \alpha_y/t=0.$ 
It is evident that for $W/t < 2.55$, all the eigenstates are delocalized as the IPR value in this region depends inversely 
on the system size across the entire energy spectrum, while all the states are localized for $W/t > 2.55.$  On the other hand, 
at the critical point the IPR spectrum behaves differently, it is neither independent of system size nor does it
scale inversely with $L$ like the extended states. This behaviour is similar to the IPR spectrum of AA model 
without RSO at the critical point, i.e. these are multifractal or in other words extended yet \textit{non-ergodic}.
Furthermore, the correlation between the energy level-spacing spectra and the scaling behaviour of IPR with L for 
all three different types of electronic states are similar to the AA model without RSO \cite{Roy-Sharma}. In the 
delocalized phase, IPR typically behaves inversely with $L$ across the whole energy-spectrum except at the special 
positions where the level-spacing jumps abruptly. At the critical point, both of them show anomalous behaviour 
with the system size, while in the localized phase IPR spectrum behaves opposite compared to the delocalized phase
at these special point of level-spacing spectra apart from being system size independent. All of these results 
are presented for PBC. OBC does not change the fundamental conclusions.   
\begin{figure*}[ht]
\centering
	\includegraphics[width=0.32\textwidth,height=0.25\textwidth]{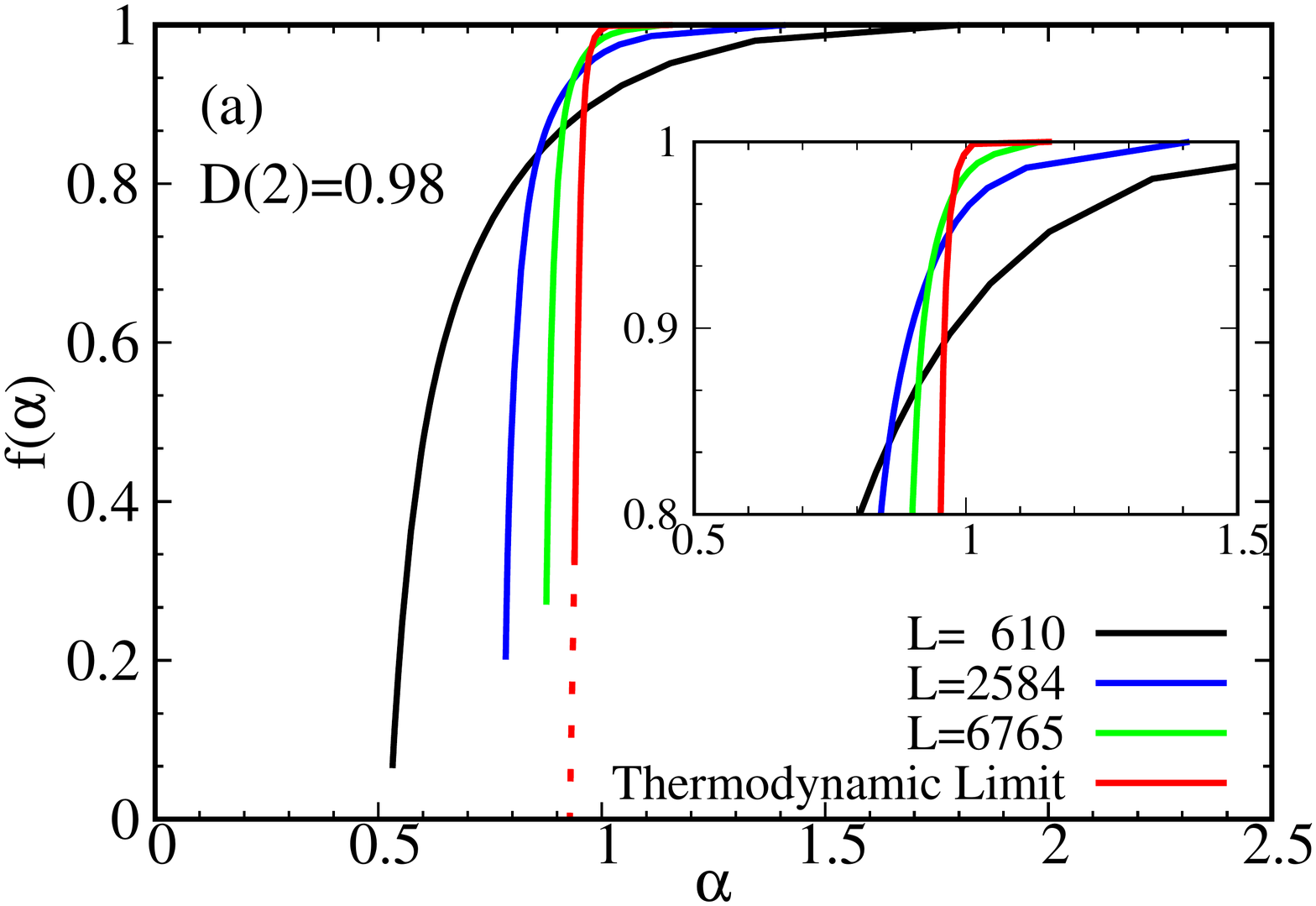}
	\includegraphics[width=0.32\textwidth,height=0.25\textwidth]{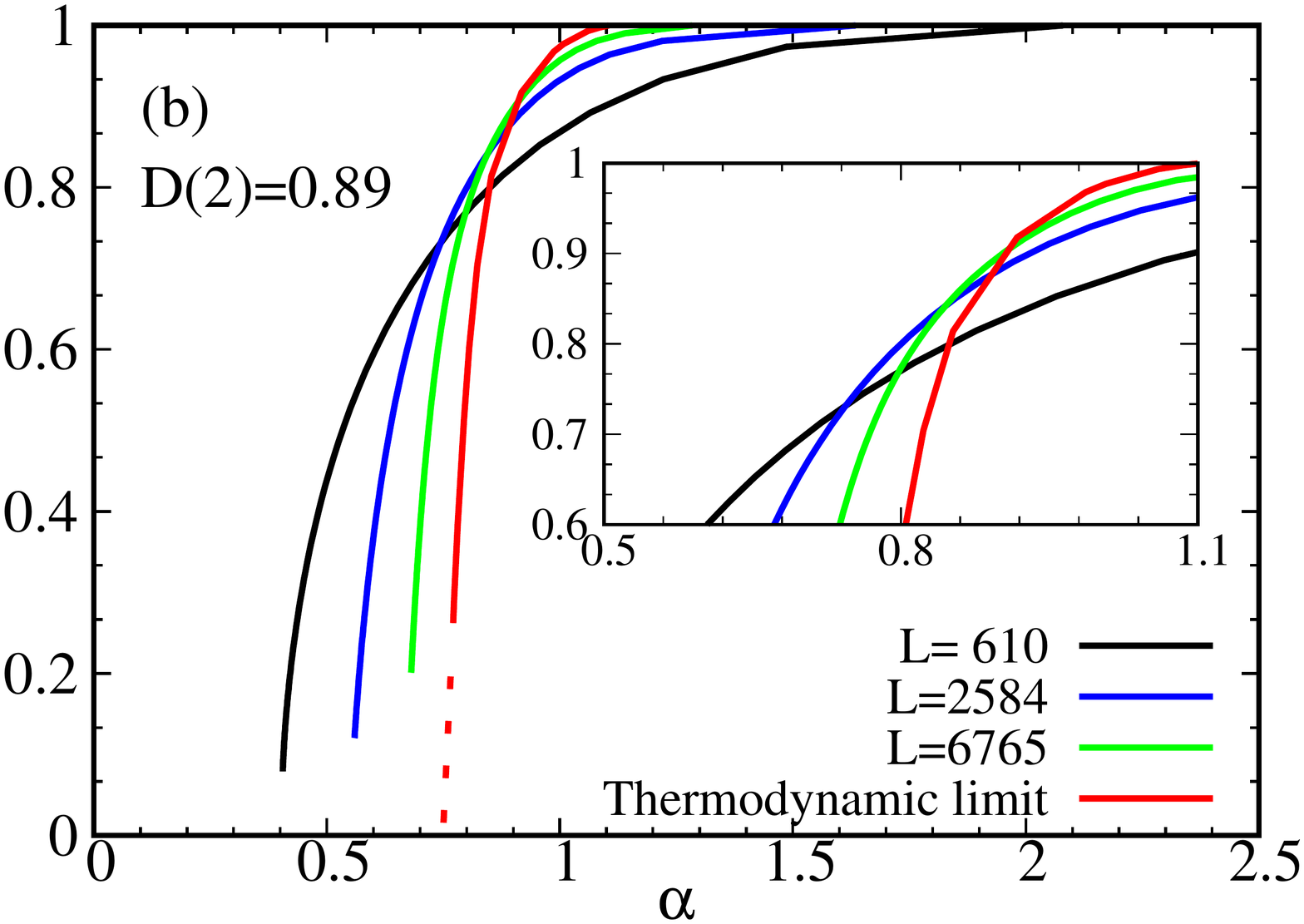}
	\includegraphics[width=0.32\textwidth,height=0.25\textwidth]{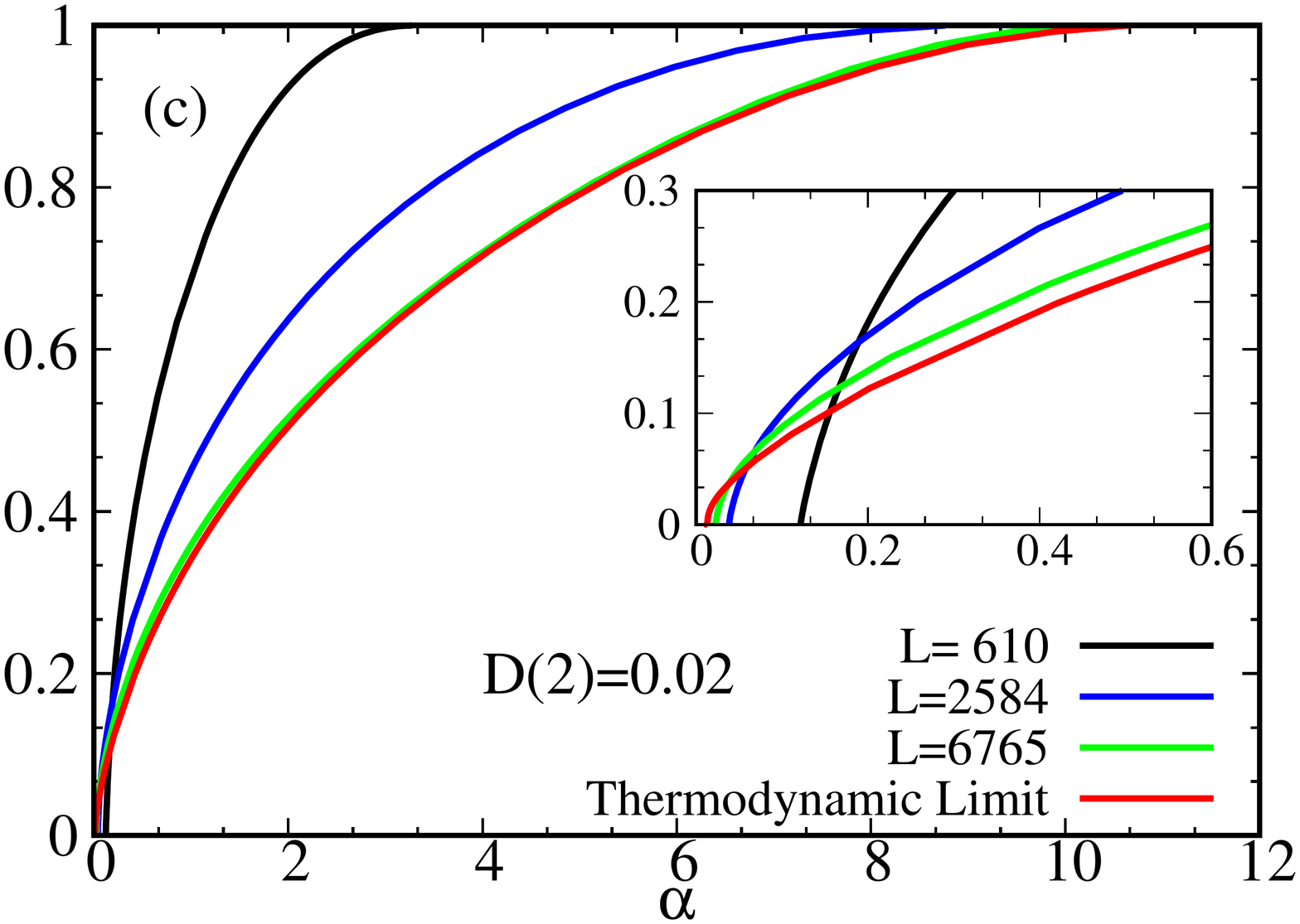} \\
	\includegraphics[width=0.32\textwidth,height=0.25\textwidth]{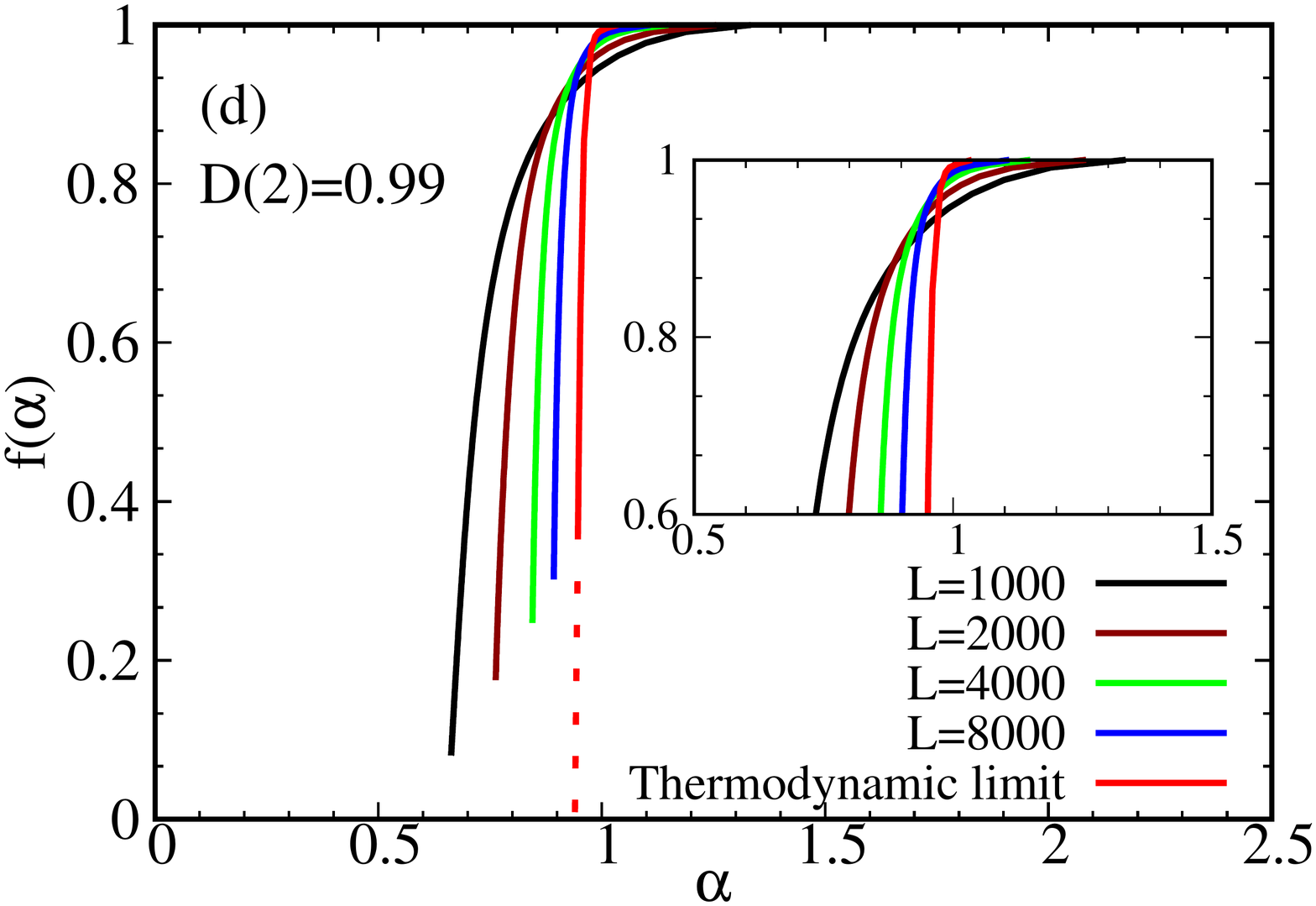}
	\includegraphics[width=0.32\textwidth,height=0.25\textwidth]{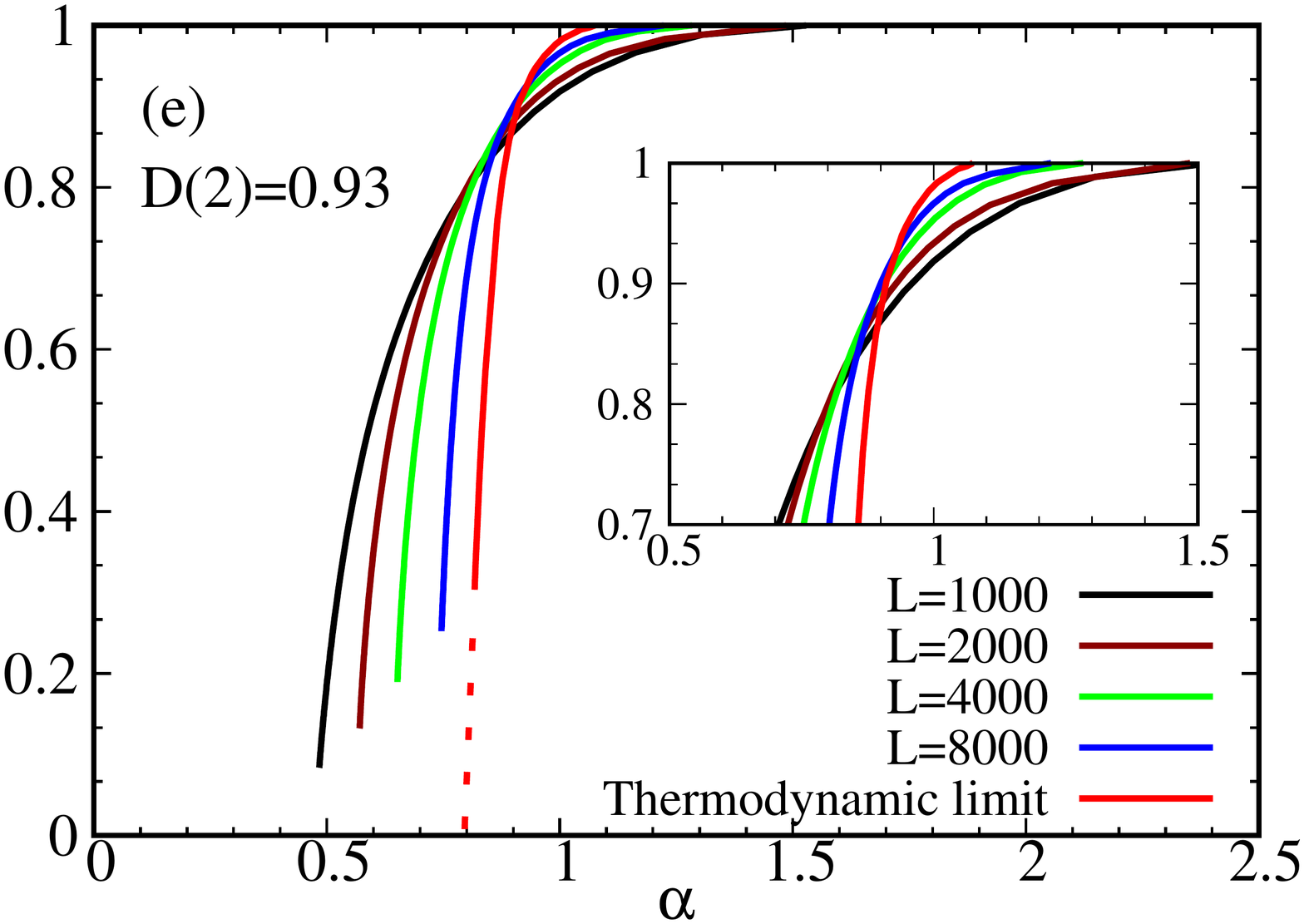}
	\includegraphics[width=0.32\textwidth,height=0.25\textwidth]{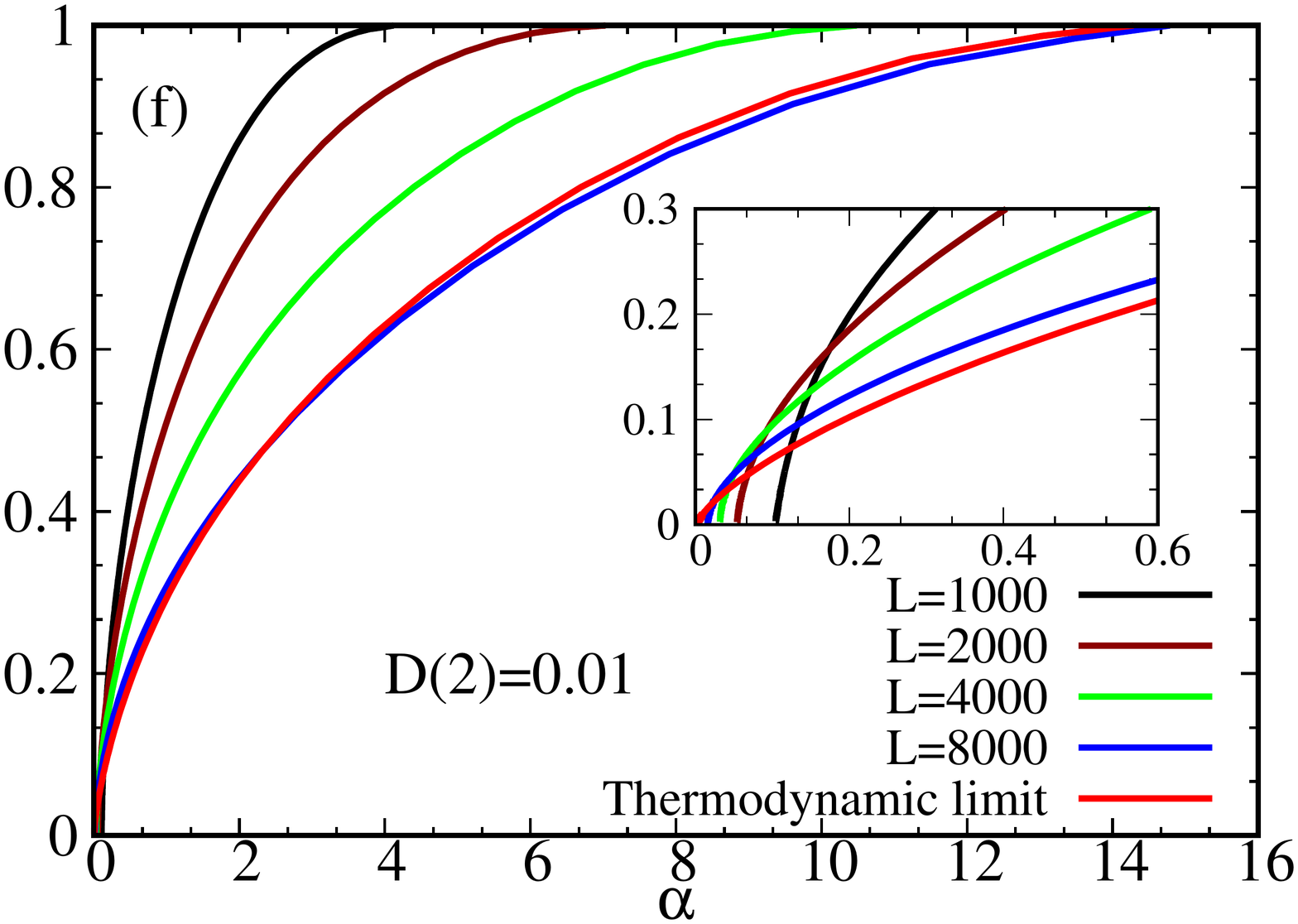}
\caption{Multifractal spectrum  of AAH Hamiltonian with RSO coupling for PBC (top row) and 
OBC (bottom row). Here $\alpha_y/t=0$ and $\alpha_z/t=0.8$. From left to right: 
(a, d)$W/t=2.5$, (b, e)$W/t=2.55$ and (c, f) $W/t=2.6$.  $D(2)$ is the generalized dimension 
$D(q)$ for $q=2$. For extended ergodic states $D(2)=1$ and $D(2)=0$ for insulating states, while 
for multifractal /non-ergodic extended states $0< D(2) < 1.$}	
\label{Fig:MFS-pbc-obc}
\end{figure*}
\section{von Neumann entropy}{\label{Sec:vNE}}
From the results of the previous sections, it is evident that at half-filling there is 
a transition from metallic phase to an insulating phase at a critical disorder strength 
$W_c/t > 2.0$, which increases as the strength of RSO is increased. Furthermore, these results also 
hint that at the critical point the states are multifractal, a preliminary observation that we are 
going to establish firmly in Sec.~\ref{Sec:multifractal-spectrum}. In this section, we present the 
results of von-Neumann entropy (vNE), an alternative indicator of single particle properties, which 
can also be used to qualititavely identify the nature of the eigenstates.  

In case of non-interacting spin-1/2 fermions, and in presence of
RSO coupling, individual eigenstates are occupied by quasiparticles.
The quasiparticle eigenstate having energy $E_{\nu}$ can be written as, 
\begin{equation}
\ket{\Psi_{\nu}} = \sum_{i=1}^{N} \left[ 
\psi^{\nu}_{i,\uparrow} \ket{1,\uparrow}_i + 
\psi^{\nu}_{i,\downarrow} \ket{1,\downarrow}_i \right], 
\end{equation}
where $\ket{1,\uparrow}_i = \ket{1}_i \otimes \ket{\uparrow}$ and 
$\ket{1,\downarrow}_i = \ket{1}_i \otimes \ket{\downarrow}$. 
$\ket{1,\uparrow}_i = c^{\dagger}_{i,\uparrow} \ket{0}$ and 
$\ket{1,\downarrow}_i = c^{\dagger}_{i,\downarrow} \ket{0}$. 
Here $\ket{0}$ represents the vacuum state for the lattice in real 
space basis. $c^{\dagger}_{i,\uparrow},~ c^{\dagger}_{i,\downarrow}$ are the 
creation operators for spin up and spin down particles respectively at the 
lattice site $i$. The average number of spin up and down particles at site $i$  
are given by $|\psi^{\nu}_{i,\uparrow}|^2 = 
\bra{\Psi_{\nu}} c^{\dagger}_{i,\uparrow} c_{i,\uparrow} \ket{\Psi_{\nu}}$ and 
$|\psi^{\nu}_{i,\downarrow}|^2 = 
\bra{\Psi_{\nu}} c^{\dagger}_{i,\downarrow} c_{i,\downarrow} \ket{\Psi_{\nu}}$ 
respectively. Then, the local density matrix $\rho^{\nu}_{j}$ can be 
obtained from the total density matrix $\rho^{\nu}$ by tracing over all the lattice sites 
except site $j$ and can then be written as,
\begin{eqnarray}
\rho^{\nu}_{j} &=&  
\left|\psi^{\nu}_{j,\uparrow} \right|^2 
\ket{1,\uparrow}_j \bra{1,\uparrow}_{j} + 
(1 - |\psi^{\nu}_{j,\uparrow}|^2) \ket{0,\uparrow}_j 
\bra{0,\uparrow}_{j} + \nonumber \\
& &
\left|\psi^{\nu}_{j,\downarrow} 
\right|^2 \ket{1,\downarrow}_j \bra{1,\downarrow}_{j} + 
(1 - |\psi^{\nu}_{j,\downarrow}|^2) \ket{0,\downarrow}_j 
\bra{0,\downarrow}_{j}
\label{Eq:site-densitymatrix}
\end{eqnarray}
It is important to note that $\ket{0,\uparrow}_j$ and $\ket{0,\uparrow}_j$ represent 
local vaccum states for $j$-th site. A similar interpretation applies to 
the states $\ket{1,\uparrow}_j$ and $\ket{1,\downarrow}_j$. The von Neumann entropy 
for spin-1/2 quasiparticles follows easily from Eq.~\ref{Eq:site-densitymatrix} as,
\begin{eqnarray}
S^{nis}_{j,\nu} &=&  
-\left(|\psi^{\nu}_{j,\uparrow} |^2 \mathrm{log}_{2}|\psi^{\nu}_{j,\uparrow}|^2  + 
(1 - |\psi^{\nu}_{j,\uparrow} |^2) \mathrm{ln}(1 - |\psi^{\nu}_{j,\uparrow} |^2)
\right)
+ \nonumber \\
& &
-\left(
|\psi^{\nu}_{j,\downarrow} |^2 \mathrm{ln}|\psi^{\nu}_{j,\downarrow}|^2  + 
(1 - |\psi^{\nu}_{j,\downarrow} |^2) \mathrm{ln}(1 - |\psi^{\nu}_{j,\downarrow} |^2) 
\right) \nonumber \\
& &
\end{eqnarray}
Finally summing over all the lattice sites, the von-Neumann entropy for a quasiparticle
eigenstate is defined as,
\begin{equation}
 S^{nis}_{\nu} = \sum_{j=1}^{N} S^{nis}_{j,\nu}.
\end{equation}
Similar to the spinless AA model, for a purely extended quasi-particle state $S^{nis}_{\nu} 
\approx (\mathrm{log}_{2} N + 1) $ and for completely localized state it is  $\approx 0$.  

In Fig.~\ref{Fig:Entropy}(a)-(c) results of our von Neumann entropy calculations are presented 
for $\alpha_z/t = 0.8$ and $\alpha_y/t=0.0$. To show the dramatic change in vNE as we move 
slightly away from the critical point ($W_c/t \simeq 2.55 $), in Fig.~\ref{Fig:Entropy}(a)
and in Fig.~\ref{Fig:Entropy}(c) we have plotted our results for $W/t=2.5$ and $W/t=2.6$
respectively. It is clear that for $W/t=2.5$ the von Neumann entropy increases with 
system size and roughly scales as expected, while it is independent of the system sizes
and the value is close to zero for $W/t=2.6$. It is clear from the results of Fig.~\ref{Fig:Entropy}(b),
at the critical point these results do not completely follow the expected pattern of purely extended or 
localized states. These results along with the localization tensor calculations and IPR data 
qualitatively capture the nature of the eigenstates at the critical point. However, 
it is necessary to have a more rigorous analysis to quantify the degree of
multifractality of the eigenstates at the critical point. To address this, in the next section,
we present a detail and careful analysis of the multifractal spectrum around the critical 
point.
\begin{figure*}[ht]
\centering
\includegraphics[width=0.32\textwidth,height=0.25\textwidth]{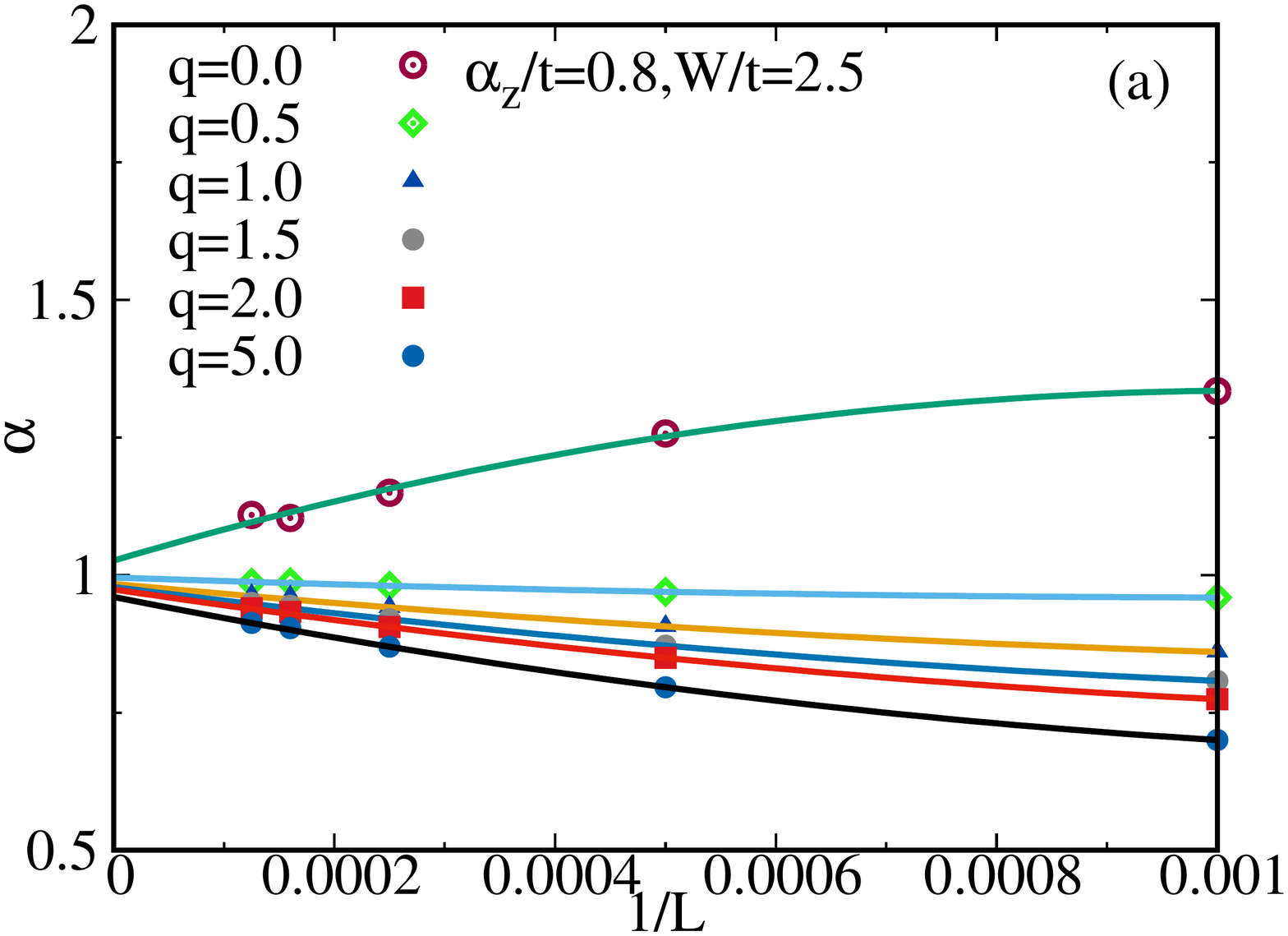}
\includegraphics[width=0.32\textwidth,height=0.25\textwidth]{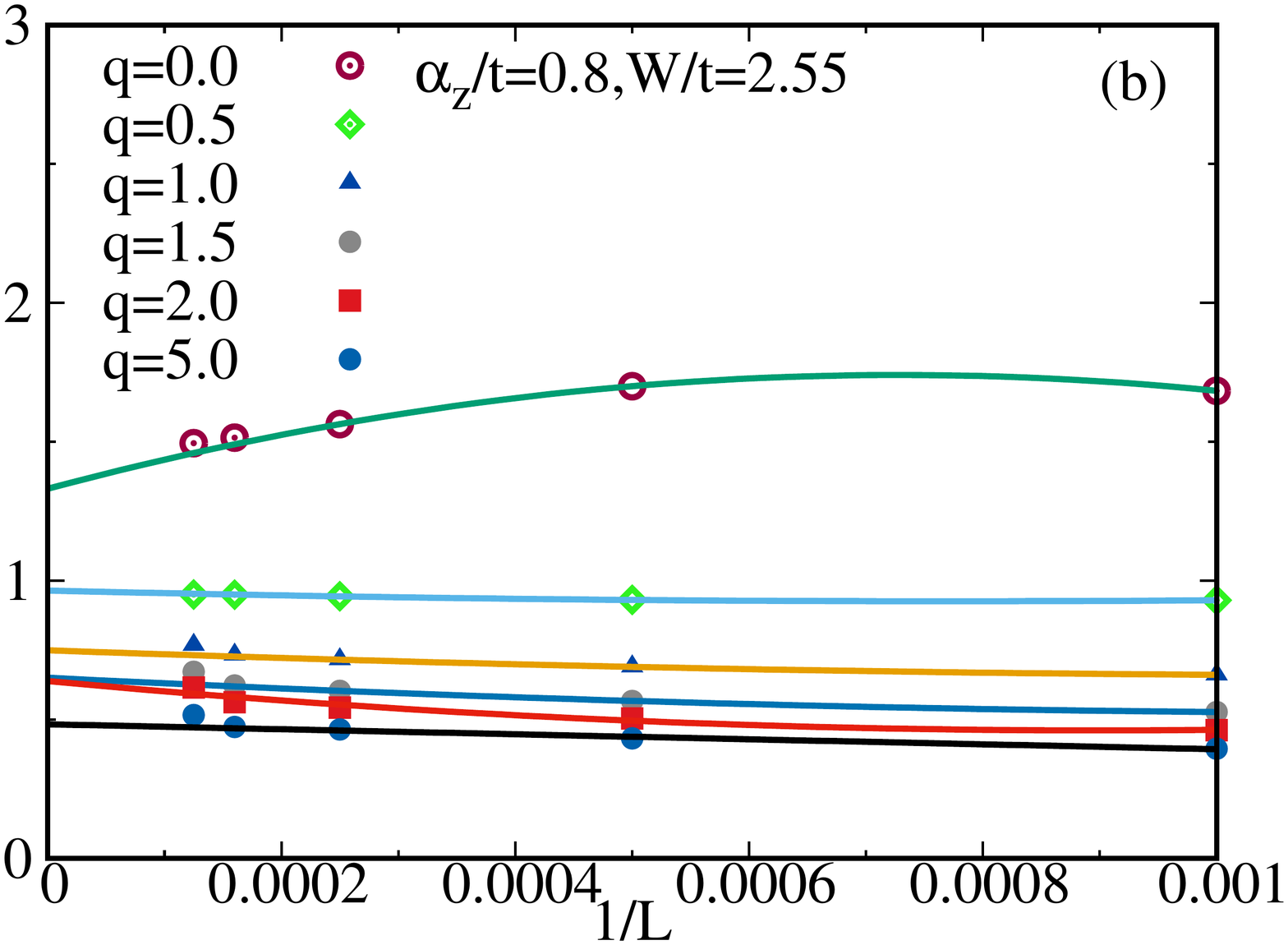}
\includegraphics[width=0.32\textwidth,height=0.25\textwidth]{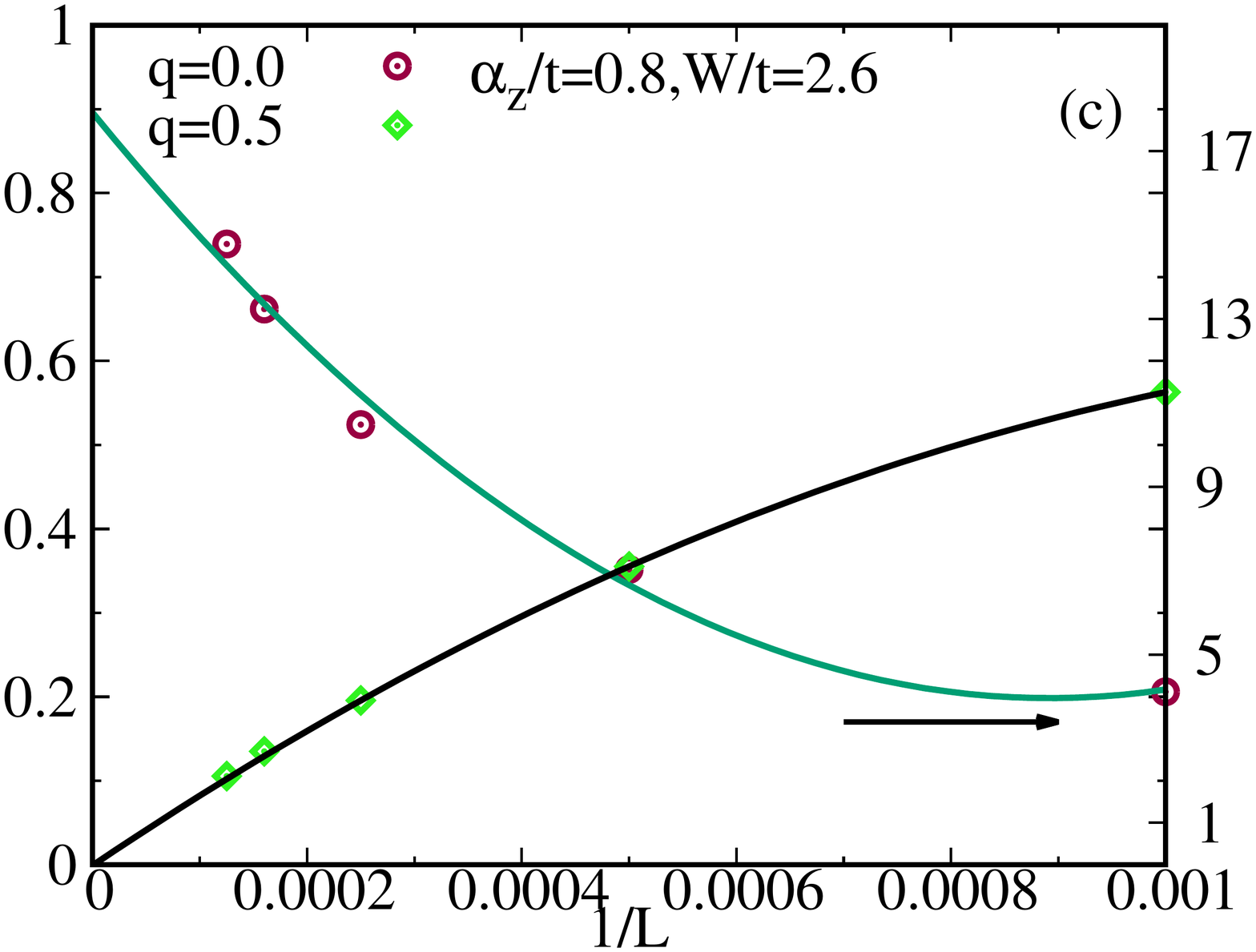}\\
\includegraphics[width=0.32\textwidth,height=0.25\textwidth]{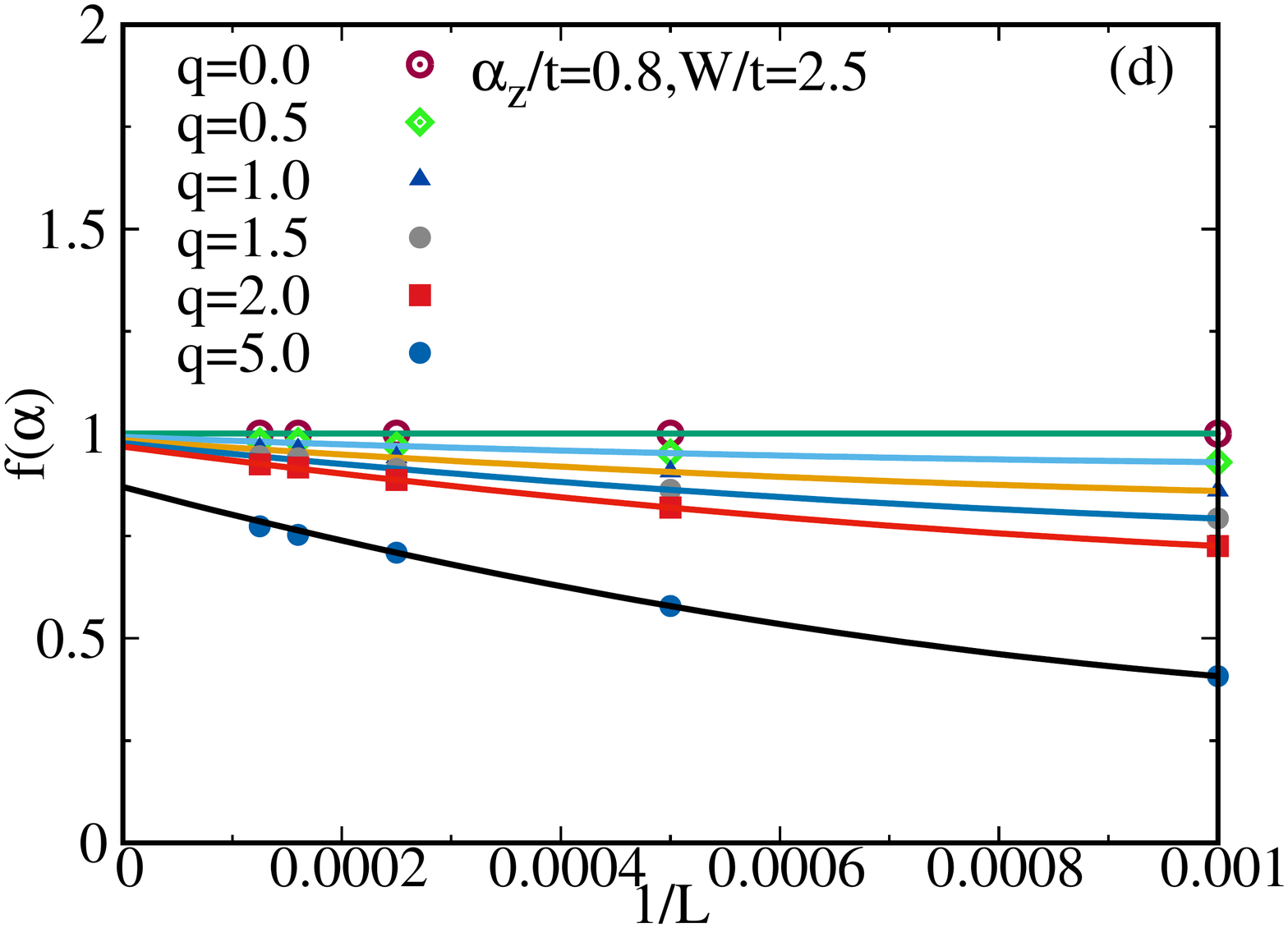}
\includegraphics[width=0.32\textwidth,height=0.25\textwidth]{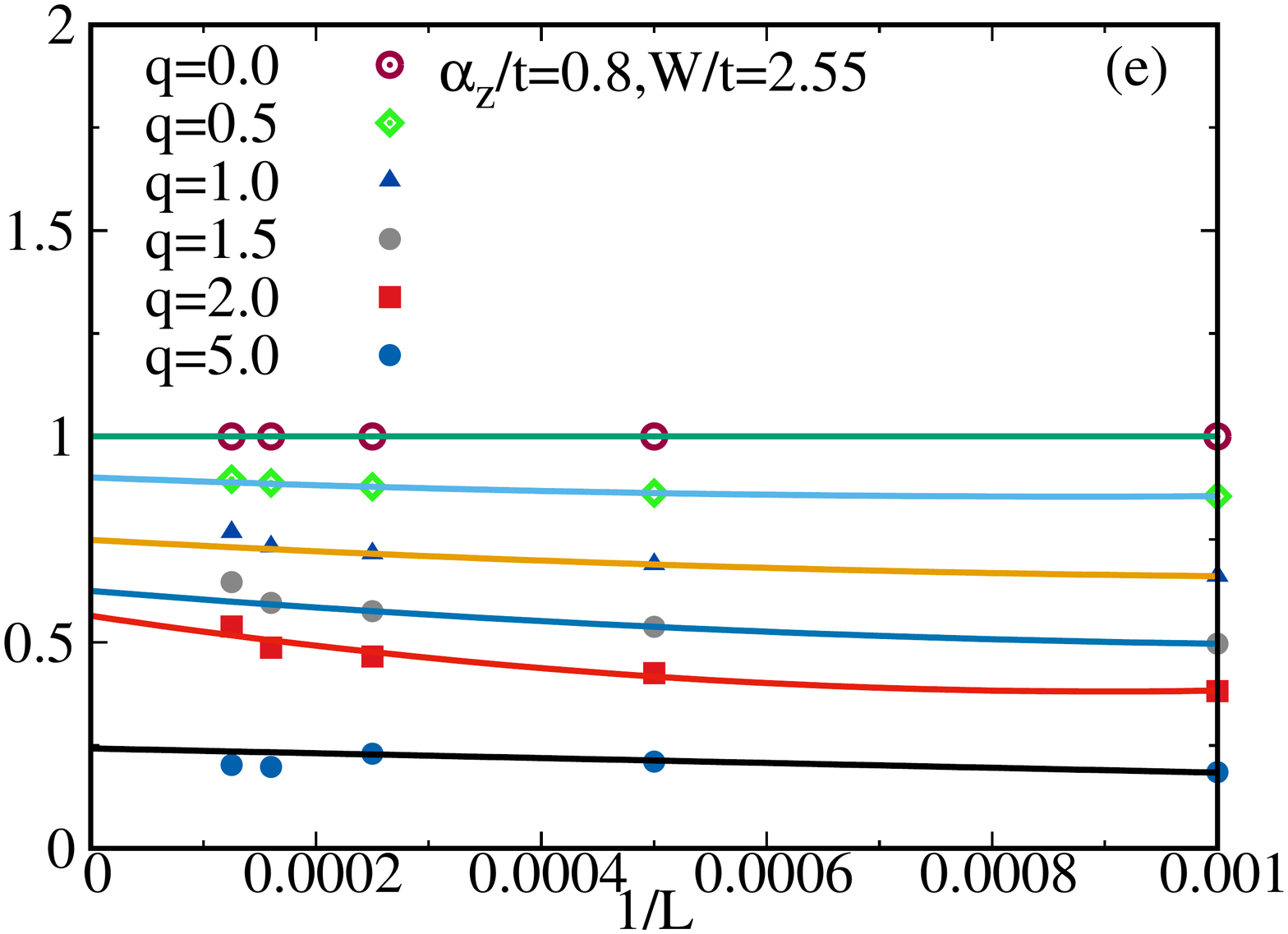}
\includegraphics[width=0.32\textwidth,height=0.25\textwidth]{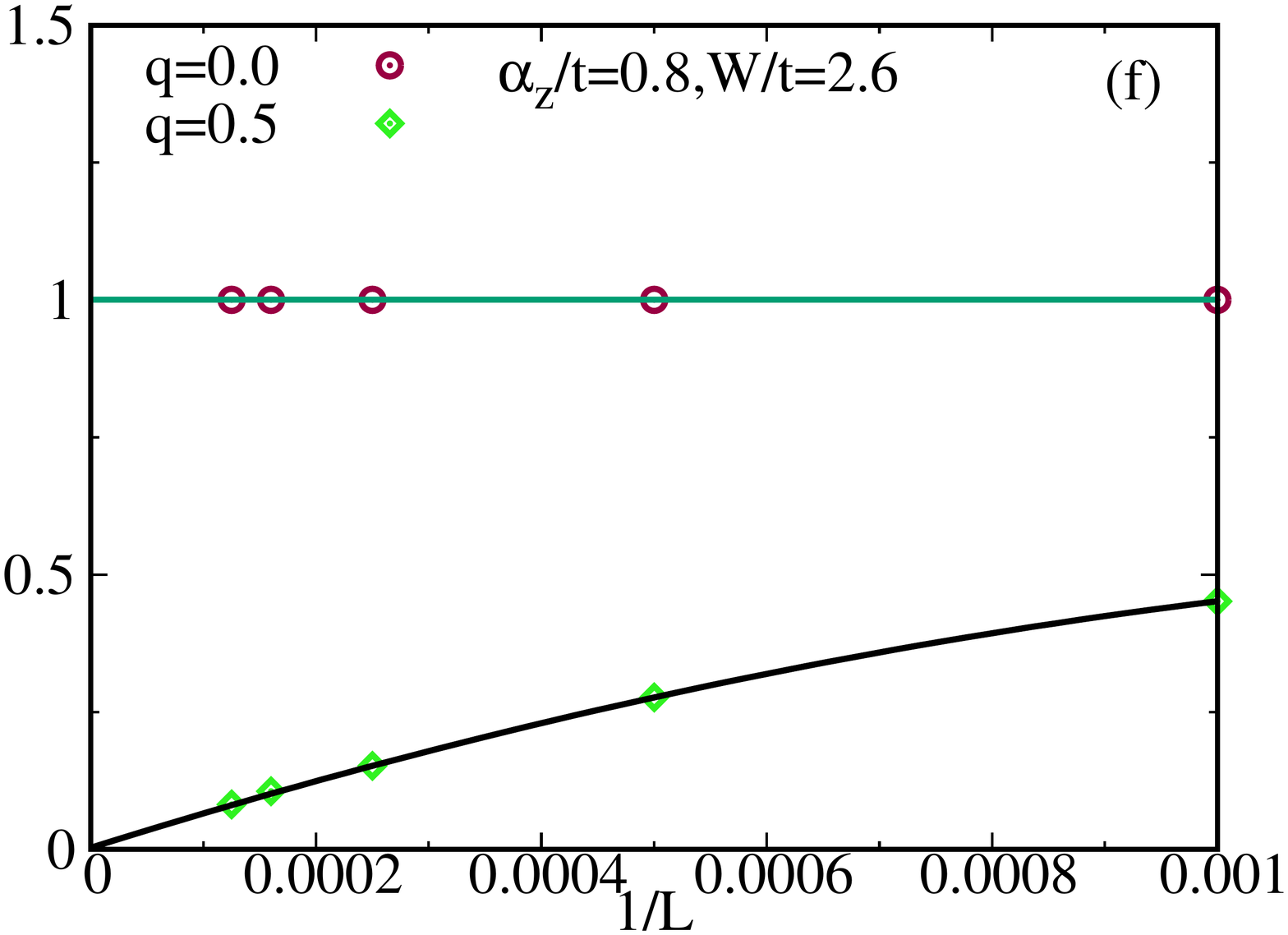}
\caption{Top row: finite size scaling of $\alpha(q)$ with $1/L$ and bottom row: finite size scaling 
of $f(\alpha(q))$ with $1/L$ for some limiting values of the moment q. The scaling functions are 
given in Eq.\ref{Eq:Scaling-Function}. The complex hopping amplitude $\alpha_y/t=0.$ From left to right,
$W/t=2.5, 2.55$, and $2.6$. All of these results are presented for open boundary condition.}	
\label{Fig:Scale-alpha-falpha}
\end{figure*}
\section{Multifractal Spectrum and the Quasi-particle Eigenstates}\label{Sec:multifractal-spectrum}
``Absence of length scale'' at the critical point of a phase transition has 
led to the understanding that localization-delocalization (LD) transition 
can also be viewed as a class of a much broader set of critical phenomena. Typically, the critical 
phenomena are characterized by critical exponents. In case of LD transition, multifractal 
analysis of the eigenstates plays a similar role like the critical exponents. Following the 
arguments of multifractal analysis \cite{De-luca} in our case, we start by identifying the  
q-th moment of the probability of finding a quasiparticle within a linear box of length 
$L~(L=Na, a=1 \textrm{~in arb. unit})$ is 
$P(q) = \sum_{i=1}^{N} \left|\psi_n(i) \right|^{2 q} \propto N^{-\tau(q)}$,
where $\psi_n$ is the quasi-particle wave function corresponding to n-th eigenvalue 
and $i=1,2,\cdots,N$. The exponent $\tau(q)$ is alternatively expressed in terms of $D(q)$ as 
\begin{equation}
\tau(q) = D(q) (q-1), 
\label{Eq:tau(q)}
\end{equation}
where $D(q)$ is called the generalized dimension. 
In case of ergodic extended (EE) eigenstates \cite{De-luca} $\tau(q) = q-1$
This conclusion follows from the argument that the real space average $P(q)/N$ converges to the 
ensemble average $\left\langle P_q\right\rangle/N = \bra{} \psi_n(i)\ket{^{2 q}}$ in the limit 
$N \rightarrow \infty$. Effectively it means that for EE states, $D(q) = 1$, whereas for a completely 
localized eigenstate $D(q) =0.$ For multifractal states, $\tau(q)$ deviates from these two limiting 
cases leading to $q$ dependence of the generalized dimension $D(q)$. These states are extended
yet \textit{non-ergodic}. Out of the possible set of generalized dimesnions, $D(2)$ is frequently 
used to characterize different states.  In practice, however, instead of computing $\tau(q)$ directly, 
an equivalent quantity $f(\alpha)$ is evaluated. It characterizes the multifractal property of the 
eigenstates. $f(\alpha)$ and $\tau(q)$ are connected by the Legendre transformation,
\begin{equation}
f(\alpha(q)) = q \alpha(q) -\tau(q), 
\label{Eq:f(alpha(q))}
\end{equation}
where $\alpha(q) = d \tau(q)/dq$. In general, $f(\alpha)$ is a smooth non-monotonic positive valued function 
having negative curvature and a global maxima but no local minima or maxima. In fact, 
$f_{max} = f(\alpha(q=0)) = d$, where $d$ is the Euclidean dimension of the system \cite{Martin Janssen}.
From the analysis of the function $f(\alpha)$, one can easily identify the nature of the eigenstates. 
For EE states, in the thermodynamic limit $f(\alpha=1)= 1$, while $f(\alpha \neq 1)= -\infty$. 
For non-ergodic (NE) eigenstates $f(\alpha(q)) \rightarrow, 0$ for $0< \alpha_{min} < \alpha(q) < \alpha_{max}$,
while $f_{\alpha(q=0)} = f_{max}$ appears for $\alpha(q=0) > 1$. In contrast to EE and NE states, for insulating 
states $f(\alpha(q)) \rightarrow 0$ as $\alpha(q) \rightarrow 0$, while 
$\alpha(q=0)$, i.e., the position of the maxima of $f(\alpha)$ spectrum shifts towards larger value than 1 
as the disorder strength is increased. It is quite evident that to identify the nature of the quasi-particle eigenstates it is sufficient to have an estimation of $\alpha_{min}$ $(f(\alpha_{min}) \rightarrow 0)$ and 
$\alpha(q=0)$ ($f_{max} = f(\alpha(q=0)) = d$). This allows us to use a well-established method 
\cite{Martin Janssen,Cuevas} to compute the multifractal spectrum for our Hamiltonian. This spectrum is computed 
and compared for PBC and OBC. Both of these boundary conditions lead to identical conclusions. 

\subsection{Calulcation of Multifractal Spectrum}{\label{Sec:MFS-Calculation}}
Before discussing the results, we briefly summarize the key steps of to compute the multifractal 
spectrum. Initially the lattice is divided into small boxes of linear size $l < L$
The first step is to find the normalized box-probability,
\begin{equation}
\mathcal{P}_{k}(l,q)=\frac{\mathcal{P}_{k}^{q}(l)}{\sum_{j=1}^{N_b}\mathcal{P}_{j}^{q}(l)},  
\end{equation}
where $1\leq k \leq N_b=L/l$ represents the $k$-th box and 
$\mathcal{P}_{k}(l,q)=\sum_{i \in l_k} \left|\psi_n(i)\right|^{2q}, l_k = l ~ \forall ~ k ,$ 
is the probability of the $n$-th eigenstate. Then $\alpha(q,L)$ and $f(\alpha(q,L))$ are  
obtained from the following relations;
\begin{eqnarray}
\alpha(q,L) &=& \underset{\delta \rightarrow 0}{\mathrm{lim}} 
\frac{\sum_{k=1}^{N_b}\mathcal{P}_{k}(l,q) \textrm{ln}(\mathcal{P}_{k}(l,1))}
{\textrm{ln} \delta}  \\
f(\alpha(q,L)) &=& \underset{\delta \rightarrow 0}{\mathrm{lim}} 
\frac{\sum_{k=1}^{N_b}\mathcal{P}_{k}(l,q) \textrm{ln}(\mathcal{P}_{k}(l,q))}
{\textrm{ln} \delta} ,
\end{eqnarray}
where $\delta = l/L$. It is important to note that this 
method of computing the multifractal spectrum is valid for $a \ll l < L$. For different 
system sizes $L$, we have chosen $l$ in a way that the above condition is satisfied and 
$0.1 \leq \delta \leq 0.5.$  $\alpha(q,L)$ and $f(\alpha(q,L))$ have been computed for 
system size upto $L=8 \times 10^3$ and averaged over the entire energy window till half 
filling. Finally, the thermodynamic limit value of 
$\alpha(q) = \underset{L \rightarrow \infty}{\textrm{lim}} \alpha(q,L)$ and 
$f(\alpha(q)) = \underset{L \rightarrow \infty}{\textrm{lim}} f(\alpha(q,L))$ 
have been estimated using finite size scaling.   
For finite size scaling, we propose the following set of functions for 
$\alpha_q(L)$ and $f(\alpha_q(L))$,
\begin{eqnarray}
\alpha_q(L) &=& \alpha_q + a_q L^{-1} + b_q L^{-2}, \nonumber \\
f(\alpha_q(L)) &=& f(\alpha_q) + c_q L^{-1} + e_q L^{-2},
\label{Eq:Scaling-Function}
\end{eqnarray}
where $\alpha_q, f(\alpha_q), a_q, b_q, c_q,$ and $e_q$ are adjustable parameters. 
In Fig.~\ref{Fig:Scale-alpha-falpha} we present results of these scaling of $\alpha_q(L)$
and $f(\alpha_q(L))$ for three different region, at the critical point and below and 
above it. Before discussing the scaling results, we first discuss the $f(\alpha(q))$ vs 
$\alpha(q)$ spectrum, presented in Fig.~\ref{Fig:MFS-pbc-obc}. In Fig.~\ref{Fig:MFS-pbc-obc}(a)-(c)
we present the results for PBC, while the results with OBC are presented in 
Fig.~\ref{Fig:MFS-pbc-obc}(d)-(f). All these results are for $\alpha_z/t = 0.8$.  As mentioned in 
earlier, one of the main purpose of presenting the results with two different boundary 
conditions is to demonstrate that it does not affect our fundamental conclusion about the nature 
of the states. Furthermore, these results indicate that one can use OBC to calculate the 
multifractal spectrum quite accurately using reasonably large system sizes, at least in 1D. 

In Fig~\ref{Fig:MFS-pbc-obc}(a) and (d),  $\alpha_z/t = 0.8$ and $W/t=2.5$. From the results 
of localization tensor, the quasi-particle states are expected to be extended and ergodic. 
It is clear from the results of multifractal spectrum that, in the thermodynamic limit 
we have $f(\alpha(q=0)) \rightarrow 1$, while $\alpha(q) \rightarrow 1$. These two values 
are not exactly $1$ as one would expect for ideal extended ergodic states, but they are 
very close to the ideal value. Using Eq.~\ref{Eq:tau(q)} and Eq.~\ref{Eq:f(alpha(q))} we have 
computed $D(2)$. As expected for metallic states, our estimated value of $D(2)$ is $0.98$
$(0.99)$ for periodic (open) boundary conditions respectively.
As we increase $W/t$ to $2.55$ (Fig.~\ref{Fig:MFS-pbc-obc} (b) and (e)), we can see the dramatic 
change in the multifractal spectrum. In the thermodynamic limit, $f(\alpha_q) \rightarrow 0$ 
for $\alpha(q) < 1$, while $f_{max}=f(\alpha(q=0))=1$ for $\alpha(q=0) > 1$. 
It clearly indicates that all the states are extended, yet \textit{non-ergodic}. This observation is 
also supported by our estimation of $D(2)$. At the critical point $D(2)=0.89$ for PBC, while $D(2)=0.93$ 
for OBC according to our estimate. 

It is interesting to have a closer look at the behaviour of $f(\alpha_q(L))$ vs. $\alpha_q(L)$ 
spectrum with system size $L$. For different system sizes the spectrum cross each other at some point. 
How these spectrum move with increasing system size on  either side of the crossing indicates 
the nature of the eigenstates. For EE and NE states, the spectrum move towards $\alpha=1$ with increasing 
system size. This evolution of the multifractal spectrum with system size gets \textit{reversed} quite 
dramatically as we increase the disorder strength just a little to the value $W/t=2.6$. In this case, 
with increasing system size the spectrum moves towards $\alpha=0$ on the the left of the crossing point, 
while on the right of the crossing point it moves further away from $\alpha=1$. The results are presented 
in Fig.~\ref{Fig:MFS-pbc-obc}(c) and (f). It is evident that in the thermodynamic limit, 
$f(\alpha_q) \rightarrow 0$ for $\alpha_q \simeq 0.0$, while  $f_{max}=f(\alpha(q=0))=1$ for 
$\alpha(q=0) >> 1$, indicating that all the states are localized. From the numerical data, we find that 
$D(2)=0.02 (0.01)$ for PBC (OBC), as expected. These results are also consistent 
with our estimation of the critical point from the localization tensor calculations, as well as 
with the IPR and vNE results. 

In Fig.~\ref{Fig:Scale-alpha-falpha}, we have presented the scaling data for  for $\alpha_y/t=0,\alpha_z/t=0.8$ and few limiting values of $q$. Here results are presented for OBC only. For PBC the results are nearly identical. 
To demonstrate the dramatic change in the multifractal spectrum we have chosen the self dual point $W_c/t=2.55$ and two different potential strengths just below and above it. In Fig.~\ref{Fig:Scale-alpha-falpha}(a)-(c), $\alpha_q(L)$ has been plotted with $1/L$, while Fig.~\ref{Fig:Scale-alpha-falpha}(d)-(f) are for $f(\alpha_q(L))$. From Fig.~\ref{Fig:Scale-alpha-falpha}(a) it is clear that just below the critical point  $\alpha_q(L) \rightarrow 1$ with increasing system size for all $q$. Same pattern can be observed for $f(\alpha_q(L))$ as well, although for higher $q$ the convergence  is not perfect. The convergence of $f(\alpha_q(L))$ to 1 becomes perfect as disorder strength is lowered slightly from $W/t=2.5$. From Fig.~\ref{Fig:Scale-alpha-falpha}(b) we can see that at the critical point $\alpha_q(L)$ and $f(\alpha_q(L))$ do not converge to a single value for different $q$ in the thermodynamic limit. $\alpha_{q=0}(L)$ converges to a value greater than 1, while it converges to a single value much less than 1 as q increases beyond 2.0. At the same time, 
$\underset{L \rightarrow \infty}{\textrm{lim}}f(\alpha_q(L))\rightarrow 0$ as $q$ is increased.
This indicates that the eigenstates are extended but non-ergodic. Quite dramatically,
as $W/t$ is increased just by a small amount to $2.6$, 
$\underset{L \rightarrow \infty}{\textrm{lim}}{\alpha_{q=0}(L)}$ converges to a value much larger than 1, 
while in the thermodynamic limit $\alpha_{q \neq 0}(L)$ and $f(\alpha_q(L))$ converge rapidly to the origin at the same time with higher moment q. This indicates that the states are localized across the entire spectrum.       
\begin{figure}[ht]
\centering
    \includegraphics[width=0.45\textwidth,height=0.35\textwidth]{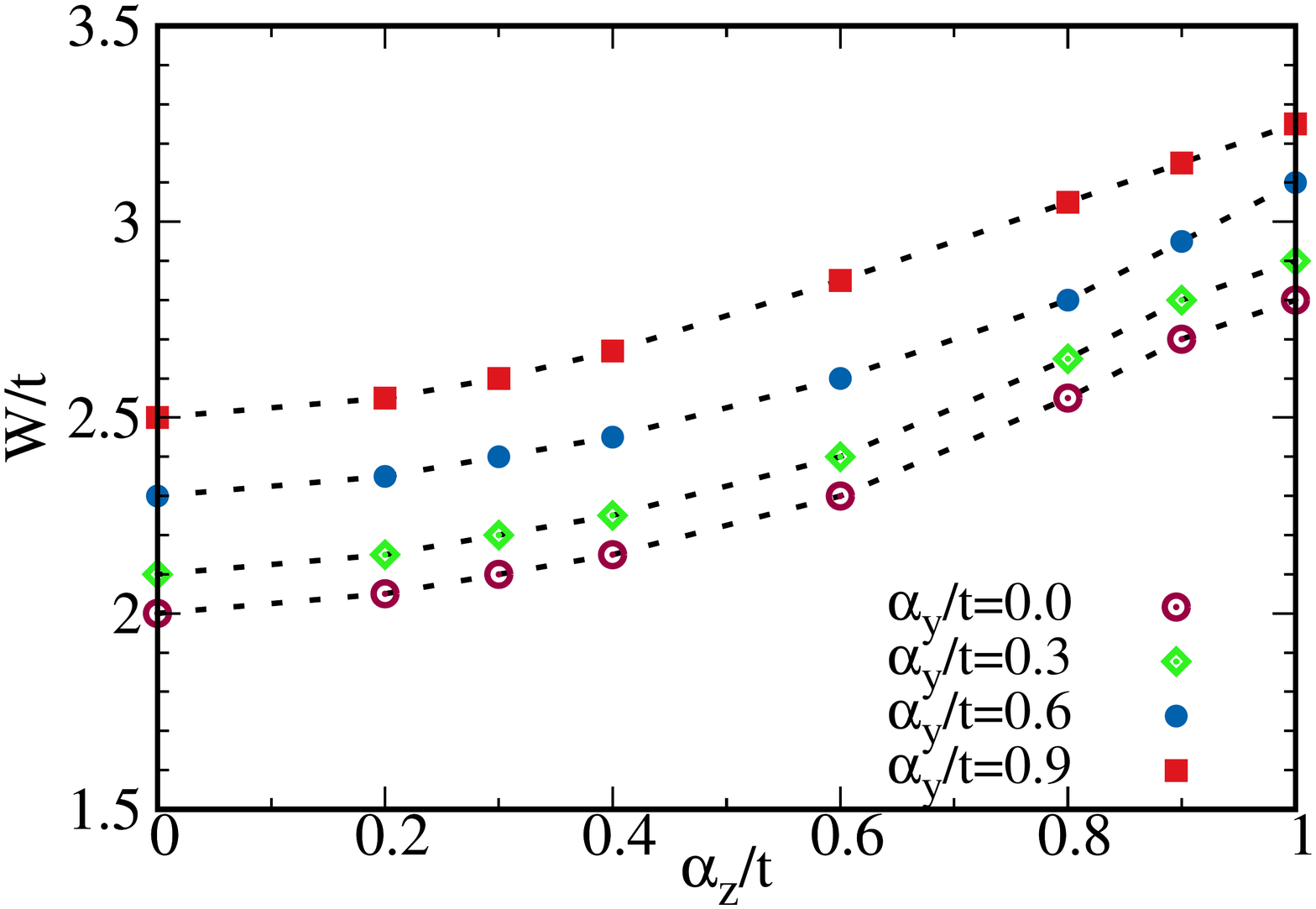}
\caption{Phase diagram in the parameter space spanned by the strength of quasi-periodic 
potential $W/t$ and the spin-flip hopping amplitude $\alpha_z/t$. $\alpha_y/t$ represents the 
spin conserving complex hopping amplitude induced by RSO coupling. The dotted lines indicate the 
phase boundary. }
\label{Fig:localization-phase}
\end{figure}
\section{Phase Diagram}\label{Sec:Phase-Diagram}
Finally, we present the phase diagram in the parameter space spanned by $W/t$ and $\alpha_z/t$. 
The phase boundaries has been obtained for increasing strength of the complex hopping of the RSO 
Hamiltonian. Each of these phase boundaries are indicated schematically by dotted line in 
Fig.~\ref{Fig:localization-phase}. The phase is metallic below it, while it is an insulating phase 
above the boundary. On each of these boundaries, all the states are extended and non-ergodic. 
\section{Conclusions}
In conclusion, we have studied the effect of RSO coupling on the critical behaviour of 
one dimensional quasi-periodic lattice, described by AAH Hamiltonian. We have found that RSO coupling 
does not affect the self-dual nature of AAH model, however the self-dual point moves towards 
higher strength of the quasi-periodic potential. Moreover, individual influences of the 
two different kinds of hopping induced by Rasha spin-orbit coupling, that is the spin conserving 
complex hopping and spin-flip real hopping, are identical in nature. We have found no evidence 
of coexistence of extended and localized states in the entire parameter space. Furthermore, the phases
are insensitive towards the existence of any sub-gap states that might exist in the energy spectrum 
depending on the lattice size and boundary conditions. In the process of studying the MIT in this system, 
we have also demonstrated that  can be also used to study the transition from delocalized to localized phase 
in system where the spin states mix and eigenstates are quasi-particles rather pure spin states. For a rigorous 
analysis of the eigenstates, we have performed the multifractal analysis, and it has been demonstrated that the 
algorithm works equally well for PBC as well OBC. This effectively removes the constraint on the system sizes that can be used to obtain the results in the thermodynamic limit for quasi-periodic systems. 
\section{Acknowledgement}
This work is supported by SERB (DST), India (File No. EMR/2015/001227, Diary No. SERB/F/6197/2016-17).
S.D would like to thank Aditi Chakrabarty for helpful feedback on the manuscript. 

\end{document}